\documentclass[twocolumn]{aastex61}
\usepackage{booktabs}
\usepackage{graphicx}
\usepackage{amsmath}

\hypersetup{
    colorlinks=true,
    linkcolor=blue,
    filecolor=magenta,      
    urlcolor=cyan,
}
\def\arcsec{\ifmmode^{\prime\prime}\;\else$^{\prime\prime}\;$\fi}
\def\arcmin{\ifmmode^{\prime}}

\received{??} %July 1, 2018
\revised{??}%June 27, 2018
\accepted{??}%\today
\submitjournal{ApJ}

\shorttitle{Mid-IR and X-Ray fit}
\shortauthors{D. Esparza}
\begin{document}
\title{Physical parameters of the torus for the type 2 Seyfert IC\,5063 \\
from mid-IR and X-ray simultaneous spectral fitting }

\correspondingauthor{D. Esparza, PhD student}
\email{d.esparza@irya.unam.mx}

\author{Donaji Esparza-Arredondo}
\author{Omaira Gonz\'alez-Mart\'in}
\affil{Instituto de Radioastronom\'ia y Astrof\'isica (IRyA-UNAM), 3-72 (Xangari), 8701, Morelia, Mexico}
\author{Deborah Dultzin}
\affil{Instituto de Astronom\'ia (IA-UNAM), Mexico city, Mexico}
\author{Cristina Ramos-Almeida}
\affil{Instituto de Astrof\'isica de Canarias (IAC), C/Vía Láctea, s/n, E-38205 LaLaguna, Spain}
\affil{Departamento de Astrof\'isica, Universidad de La Laguna (ULL), E-38205 La Laguna, Spain}
\author{Jacopo Fritz}
\affil{Instituto de Radioastronom\'ia y Astrof\'isica (IRyA-UNAM), 3-72 (Xangari), 8701, Morelia, Mexico}
\author{Josefa Masegosa}
\affil{Instituto de Astrof\'isica de Andaluc\'ia (CSIC), Glorieta de la Astronom\'ia s/n 18008, Granda, Spain}
\author{Alice Pasetto}
\affil{Instituto de Radioastronom\'ia y Astrof\'isica (IRyA-UNAM), 3-72 (Xangari), 8701, Morelia, Mexico}
\author{Mariela Mart\'inez-Paredes}
\affil{Korea Astronomy and Space Science Institute 776, Daedeokdae-ro, Yuseong-gu, Daejeon 34055, Republic of Korea}
\author{Natalia Osorio-Clavijo}
\affil{Instituto de Radioastronom\'ia y Astrof\'isica (IRyA-UNAM), 3-72 (Xangari), 8701, Morelia, Mexico}
\author{Cesar Victoria-Ceballos}
\affil{Instituto de Radioastronom\'ia y Astrof\'isica (IRyA-UNAM), 3-72 (Xangari), 8701, Morelia, Mexico}

\begin{abstract}
%To understand the diversity of classes observed in active galactic nuclei (AGN), a geometrically and optically thick torus of gas and dust is required to obscure the central engine depending on the line of sight to the observer.
In order to understand the diversity of classes observed in active galactic nuclei (AGN), a geometrically and optically thick torus of gas and dust is required to obscure the central engine depending on the line of sight to the observer. 
%its inner parts for some lines of sight. However, this torus is not spatially resolved even for the closest AGN.
%Spectroscopic studies have been broadly used to characterize the main properties of the torus. Nevertheless, the torus models are described by a large number of parameters that have not been \textbf{constrained} completely yet. 
We perform a simultaneous fitting of X-ray and mid-infrared (mid-IR) spectra to investigate if the same structure could produce both emissions and, if this the case, to obtain better constraints for the physical parameters of the torus. In this case we take advantage of the fact that both emissions show important signatures of obscuration.
%\textbf{Nevertheless}, the torus has a large number of parameters that have not been constrained yet. Here, we study the combination of X-ray and mid-infrared spectra to better constrain the physical parameters of the torus taking advantage \textbf{of the fact} that both show important signatures of obscuration.
We used the nearby type-2 active nucleus IC\,5063 as a test object. This object is ideal because of the wealth of archival data including some high resolution data. It also has a relatively high AGN luminosity that dominates at both X-ray and mid-IR frequencies. We use high spectral resolution \emph{NuSTAR} and IRS/\emph{Spitzer} spectra. The AGN dusty models used several physically motivated models. 
%for mid-IR are: the smooth torus \citep{Fritz06,Feltre12}, the clumpy torus \citep{NenkovaA08,NenkovaB08}, and CAT3D-wind \citep{Hoenig17}. For the X-ray reflection component we used the borus02 model described by \citet{Balokovic18}. We constructed several baseline models, combining X-rays and mid-IR models. 
%These baselines models were used to simultaneously fit mid-IR and X-ray data.
We found that the combination of the smooth torus models at mid-IR by \citet{Fritz06} and at X-rays by \citet{Balokovic18}, with the viewing and half-opening angles linked to the same value, is the best choice to fit the spectra at both wavelengths. This allows us to determine all the parameters of its torus. This result suggests that the structure producing the continuum emission at mid-IR and the reflection component at X-ray is the same.
%This result confirms for the first time a scenario where the structure that produces the continuum emission at mid-IR and the reflection component at X-ray is the same. 
Therefore, we prove that this technique can be used to infer the physical properties of the torus, at least when AGN dust dominates the mid-IR emission and the reflection component is significant at X-rays.
%We used these mid-IR models and the borus02 model described by \citet{Balokovic18} to simultaneously fit mid-IR and X-ray data.
%We used the Clumpy models described by \citet{NenkovaA08,NenkovaB08} for mid-IR spectra and the radiative transfer code Borus described by \citet{Balokovic18} for X-ray spectra. Borus model can be fitted within the X-ray spectral fitting software XSPEC. We develop a code able to convert the mid-IR models and IRS/\emph{Spitzer} spectra into XSPEC format to simultaneously fit mid-IR and X-ray data. mid-IR data are able to constrain ? parameters. Parameter ? is degenerated. X-ray spectral alone are able to constrain ? parameters. Parameter ? are degenerated. The simultaneous fit is able to better constrain ? parameters, as demonstrated by the test perform on the \textbf{degeneracy} of the parameters.
%Thus, this technique can be used to infer the physical properties of the torus. IC\,5063 shows a torus with ?.
\end{abstract}

%%%%%%%%%%%%%%%%%%%%%%%%%%%%%%%%%%%%%%%%%%%%%%%%%%%%%%%%%%
%%%%%%%%%%%%%%%%%%%%%%%%%%%%%%%%%%%%%%%%%%%%%%%%%%%%%%%%%%
%%%%%%%%%%%%%%%%%%%%%%%%%%%%%%%%%%%%%%%%%%%%%%%%%%%%%%%%%
\section{Introduction}\label{sec:intro}

According to the simple unification model (UM) of active galactic nuclei (AGN), a toroidal structure (broadly referred as the torus) provides the anisotropic obscuration needed to explain the diversity of AGN properties observed across the electromagnetic spectrum \citep{Antonucci93,Urry95}. 
%More recent works suggest that this torus actually changes for different luminosities and/or accretion rates \citep{Shlosman05,Elitzur16}. 
The line of sight (LOS) to the observer with respect to the torus, its geometry, chemical composition, and distribution are keys to understand AGN diversity, perhaps linked to fundamental changes for different AGN classes \citep{Shlosman05,Elitzur16}. 

This torus absorbs optical/UV accretion disc radiation and re-emits it at infrared wavelengths \citep[see][for a review]{Netzer15,Ramos-Almeida17}. Radiative transfer models based dust distributed on a toroidal geometry have been proven to be successful in reproducing the infrared spectral energy distribution (SED) of AGN \citep[e.g.][]{Hatziminaoglou08,Hatziminaoglou09,Ramos-Almeida09,Alonso-Herrero11,Hoenig17}. Initially, most authors used smooth dust distributions with different radial and vertical density profiles by simplicity \citep[e.g.][]{Pier93,Granato94,Efstathiou95,Schartmann05,Fritz06}. It was later proposed that the dust is most probably arranged in clouds instead of being smoothly distributed \citep[e.g.][]{Krolik88,Tacconi94}. The dusty torus has been the subject of several kinds of models that aimed to extract physical (e.g. optical depth) and geometrical (e.g. orientation and size) properties from SED and, in some cases, interferometric observations. We can divide them into four kinds: smooth \citep{Fritz06}, clumpy \citep{NenkovaA08,NenkovaB08,Hoenig10a,Hoenig10b}, smooth + clumpy \citep{Stalevski12,Siebenmorgen15}, and windy \citep{Hoenig17}. Since two decades already, adjusting models to IR spectra and broad band SEDs, has been the one and only mean in the attempt to derive clues on the dust geometry, composition and on its geometrical distribution.
%\textbf{Thus, modeling the infrared} AGN spectra provides evidence of the dust geometry, composition and distribution. 

Furthermore, signatures of reprocessing emission by the torus in the X-ray band arise primarily from interaction of X-ray photons with the surrounding gas \citep{Ghisellini94,Krolik94}. The main two features are the neutral iron line around 6.4 keV (FeK$\rm{\alpha}$) and the Compton hump peaking at $\rm{\sim}$10-30 keV. These features have been observed in the X-ray spectra of most AGN \citep[e.g.][]{Matt91, Ricci14}. Reprocessed continua are known to vary as a function of geometry of the reprocessing material \citep{Nandra94}. It has been suggested that both the Compton hump and the narrow cores of the FeK$\rm{\alpha}$ emission line in AGNs are likely produced in the torus \citep[see][]{Shu10,Liu10,Fukazawa16}, being an ubiquitous component in Seyfert galaxies \citep{Bianchi04}. Therefore, X-ray spectral fitting to the high energy continuum emission (above 10 keV) and the FeK$\rm{\alpha}$ line might provide important information about the torus geometry, cloud distribution, and opacity.

%\textbf{Indeed, recently significant FeK$\rm{\alpha}$ line variability of several tens of percent has been detected with timescales larger by a factor of 10-100 than the inner radius of the torus, consistent with the idea that the main origin for a narrow FeK$\rm{\alpha}$ line comes from the X-ray reflection by a torus \citep{Fukazawa16}. Additionally, the FeK$\rm{\alpha}$ line has been claimed to show anisotropic emission, which links its origin to the torus \citep{Liu10}.}
%Narrow neutral FeK$\rm{\alpha}$ lines are confirmed to be an ubiquitous component in Seyfert spectra \citep{Bianchi04}. 
%The scenario in which this narrow line is originated in the inner parts is consistent with the UM and suggest the presence of the torus in (almost) all sources, even if unobscured. 

%Indeed, recently significant FeK$\rm{\alpha}$ line variability of several tens of percent was detected with timescales larger by a factor of 10-100 than

The 100-fold increase in sensitivity in the hard X-ray band ($\rm{>}$10 keV) brought by \emph{NuSTAR} \citep{Harrison13} made possible to study the spectral signatures of the torus for the first time. Empirically, spectral models with approximately toroidal geometry have been calculated by \citet{Murphy09} (MyTorus), \citet{Ikeda09}, \citet{Brightman11} (BNtorus), \citet{Liu15} (ctorus), \citet{Furui16} (MONACO), and \citet{Balokovic18} (borus02). Several of them are currently available to the community. %Recently,  present a new grid of X-ray spectral templates based on radiative transfer calculations in neutral gas in an approximately toroidal geometry. The main advantage of this model is that the main parameters are closely linked to the Smooth torus models.

For this work, we selected the type-2 Seyfert IC\,5063 as a test object. This AGN is located at the center of a nearly lenticular galaxy at 46\,Mpc \citep[][]{Alonso-Herrero11}.
%\citep[z = 0.0113,][]{Morganti98, Morganti07,Alonso-Herrero11}.
This galaxy contains a disc with large-scale dust lanes \citep[][]{Morganti98}, possibly resulting from a merger \citep[][]{Morganti98}. According to \citet{Ichikawa15} the bolometric luminosity of IC\,5063 is $\rm{3.38 \times 10^{44}\, erg \,s^{-1}}$.
At X-ray wavelengths, IC\,5063 has been observed with \emph{GINGA} \citep{Koyama92}, \emph{ASCA} \citep{Tanaka94}, and \emph{ROSAT} \citep{Pfeffermann87,Vignali97}. More recently, IC\,5063 was observed at X-ray wavelengths with \emph{NuSTAR} \citep[][see also Balokovic in prep for details on the X-ray spectral analyses]{Balokovic18}.
At infrared wavelengths, \citet{Peeters04} observed this source with the ISO satellite determining that it is dominated by the AGN with little evidence of polycyclic aromatic hydrocarbon (PAHs) molecule emission. The dusty torus properties of IC\,5063 have been explored through high angular resolution near and mid-IR photometric data \citep{Ramos-Almeida09, Ramos-Almeida11, Alonso-Herrero11} and clumpy models of \citet{NenkovaA08,NenkovaB08}, allowing a direct comparison with our results.
%\citet{Vignali97} reported the $\rm{0.1-10}$\,keV spectrum of IC\,5063 using \emph{ASCA} and \emph{ROSAT}. They used two power-law components (with photon indices of $\rm{1.7 \pm 0.2}$ and $\rm{2.2 \pm 0.3}$, where the former is absorbed with hydrogen column density ($\rm{N_H}$) $\rm{\sim 2 \, \times 10^{23} \, cm^{-2}}$ \citep[][]{Tazaki11}). 
%us \textbf{to compare them directly with our results.} %direct comparison with our results.

In this paper, we present a new technique to combine X-ray and mid-IR spectral information to make a simultaneous fit to torus models. We demonstrate that our method can successfully constrain the torus parameters and obtain information more complete in both ranges of wavelengths. %, that are not accessible using only X-ray or mid-IR spectra. 
The X-ray and mid-IR observations are presented in Section \ref{sec:Data}. Subsequently, the spectral fitting methodology is shown in Section \ref{sec:fitting}. The main results and discussion, within the framework of our goals, are presented in Sections \ref{sec:resultsfit} and \ref{sec:Discussion}. Finally, a brief summary and conclusions are given in Section \ref{sec:Conclusion}.

%%%%%%%%%%%%%%%%%%%%%%%%%%%%%%%%%%%%%%%%%%%%%%%%%%%%%%%%%%%%%%%%%
%%%(1)-> dca remove Sept 18: In the \emph{Suzaku} and \emph{NuSTAR} archives, there is only one available observation covering energies above 10 keV. Since the \emph{Suzaku} data covers energies below 2 keV, they have not been used for this study. However,
\section{Data}\label{sec:Data}
\subsection{X-ray data}\label{subsec:dataX-ray}

There are several X-ray observations available in the archives of different satellites for IC\,5063. However, we need to cover energies above 10 keV because it is the aim of this paper to constrain the reflection component associated to the torus. %%%(1)
The \emph{NuSTAR} is the first focusing hard X-ray telescope with high sensitivity\footnote{https://heasarc.gsfc.nasa.gov/docs/nustar/}. This gives the advantage to observe with a single mode from $\rm{\sim}$3-79\,keV, perfectly suited to study the AGN reflection component.
Therefore, we use the hard band spectrum observed with \emph{NuSTAR} \citep{Harrison13}, including both FPMA and FPMB focal plane modules. \emph{NuSTAR} has observed IC\,5063 once (ObsID 60061302002, P.I. Harrison) on July 8th of 2013.
%% SUGERENCIA ALICE: In the Suzaku and NuSTAR archives, there are one available observation of the source of interest. As explained before, the suitable data for this kind of study are those covering energies above 10 keV. Since the Suzaku data covers energies below 2 keV, they have not beed used for this study. Instead, we used the Nustar data which covers the wide range between 2 to 79 keV.
%The availability of two \emph{NuSTAR} observations will allow us to use absorption variations to constrain the torus properties, as suggested by \citet{Balokovic18}. 
% The satellite has a field of view 13 x 13 arcsec.

\emph{NuSTAR} data reduction was done using the data analysis software \emph{NuSTARDAS} v.1.4.4 distributed by the High Energy Astrophysics Archive Research Center (HEASARC). The calibrated, cleaned and screened event files were generated using the {\sc nupipeline} task (CALDB 20160502). A circular region of 1 arcmin radius was taken to extract the source and background spectrum on the same detector and to compute the response files (RMF and ARF files) using the {\sc nuproducts} package within \emph{NuSTARDAS}. Finally, we used the {\sc grppha} task within the FTOOLS to group the spectra with at least 60 counts per bin. The net exposure is 18.4 ksec. We found some cross-calibration issues between the FPMA and FPMB modules, larger below $\rm{\sim}$3 keV. We used the \emph{NuSTAR} data above 3 keV to avoid them.

\subsection{Mid-IR data}\label{subsec:dataMid-IR}
\begin{figure}
    \centering
    \includegraphics[scale=0.22]{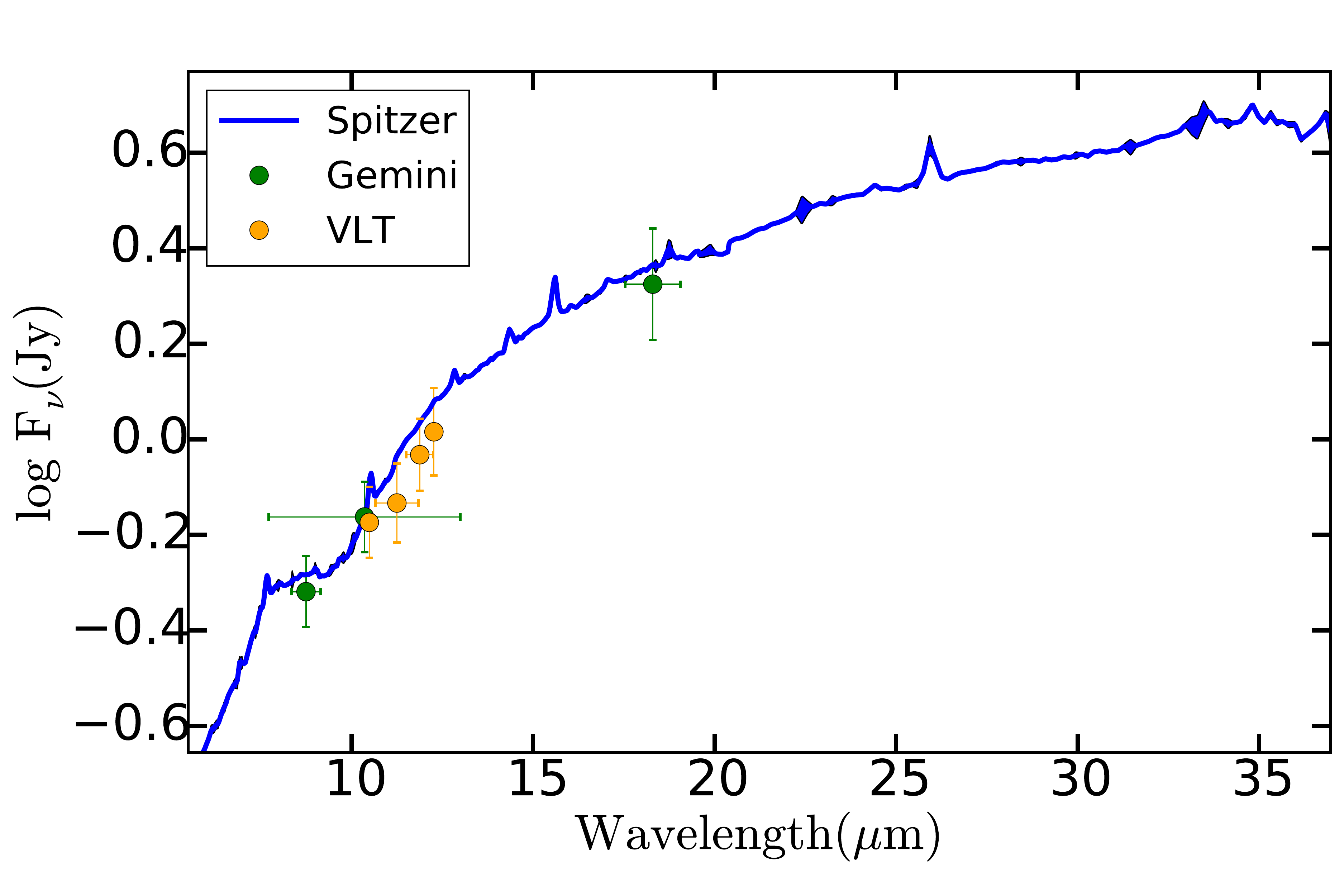}
    \caption{The \emph{Spitzer}/IRS spectrum and photometric data. The orange and green points are measurements from VLT and Gemini, respectively.} %\textbf{The flux error bars includes $\rm{15\%}$ of the flux as flux calibration error \citep[see][]{Ramos-Almeida09}.}}
    \label{fig:mid-IRSpec}
\end{figure}
Regarding the IR data, we used the high-resolution \emph{Spitzer}/IRS spectrum downloaded from the CASSIS\footnote{http://cassis.astro.cornell.edu/atlas/} catalog \citep[the Cornell AtlaS of \emph{Spitzer}/IRS Sources,][]{Lebouteiller11}. The spectral resolution of \emph{Spitzer}/IRS ($\rm{R \sim 60-130}$) is similar to that obtained by ground based observations. The \emph{Spitzer}/IRS spectrum could have a high contribution of galaxy emission due to its relative low spatial resolution. Note that we could overcome this issue by including stellar libraries to the fit. However, the inclusion of these libraries significantly worsen the estimate of the resulting parameters \citep{Gonzalez-Martin19a,Gonzalez-Martin19b}. In order to investigate it, we compared this spectrum with ground-based high spatial resolution fluxes (see Figure \ref{fig:mid-IRSpec}). In particular, we compiled VISIR/VLT and T-ReCS/Gemini fluxes in Si2$\rm{_{8.73\mu m}}$, N-band$\rm{_{10.4\mu m}}$, SIV$\rm{_{10.5\mu m}}$, PAH2$\rm{_{11.3\mu m}}$, NEII\_1$\rm{_{12.3\mu m}}$, and Qa$\rm{_{18.2\mu m}}$ filters reported in NED\footnote{https://ned.ipac.caltech.edu/} \citep{Hoenig10a, Ramos-Almeida11}. These data provide high-spatial resolution fluxes ($\sim 100$\,pc). 
%, we compared this spectrum with ground-based high spatial resolution images. \textbf{In particular, we compiled VISIR/VLT and T-ReCS/Gemini fluxes in Si2$\rm{_{8.73\mu m}}$, N-band$\rm{_{10.4\mu m}}$, SIV$\rm{_{10.5\mu m}}$, PAH2$\rm{_{11.3\mu m}}$, NEII\_1$\rm{_{12.3\mu m}}$, and Qa$\rm{_{18.2\mu m}}$ filters reported in NED\footnote{https://ned.ipac.caltech.edu/} \citep{Hoenig10a, Ramos-Almeida11}.} These data provide high-spatial resolution fluxes ($\sim 100$\,pc). 
%Figure \ref{fig:mid-IRSpec} compares the high spatial resolution photometric data with the \emph{Spitzer}/IRS spectrum.
We found that the \emph{Spitzer}/IRS spectrum shows slightly higher fluxes than the VLT and Gemini photometric data points, although those are well in agreement when ground-based flux calibration uncertainties are taken into account \citep[15\% at N-band and 25\% at Q-band of the flux according e.g.][]{Ramos-Almeida11}.

In fact, \citet{Asmus14} studied the T-ReCS and VISIR images of IC\,5063 and found a compact but consistently elongated mid-IR nucleus (FWHM(major axis)$\sim$ 0.52 arcsec $\sim$ 110\,pc; PA$\sim$ 107$\rm{^{\circ}}$) without any further host emission detected. They found that the direction of this elongation coincides with the extended [O\,III] line emission. Additionally, \citet{Hoenig10a} compares the photometric data with the \emph{Spitzer}/IRS and VISIR spectra extracted over 0.75 arcsec and found that the that agrees well. Indeed, according to \citet{Panuzzo11} the continuum of IC\,5063 in the low resolution \emph{Spitzer}/IRS spectrum is dominated by hot dust, most probably coming from the AGN torus. They did not find PAH feature emission, although some forbidden lines were detected. The lack of strong stellar or starburst components makes the \emph{Spitzer/}IRS spectrum of IC\,5063 ideal to study the torus parameters throughout mid-IR spectral fitting.
%Therefore, the \emph{Spitzer}/IRS spectrum does not \textbf{greater} show any signs of strong nuclear starburst.

In order to perform spectral fitting to the data, we converted the mid-IR \emph{Spitzer/}IRS spectrum into XSPEC format using {\sc flx2xsp} task within HEASOFT. This tool reads a text file containing one or more spectra and errors and writes out a standard XSPEC pulse height amplitude (PHA\footnote{Engineering unit describing the integrated charge per pixel from an event recorded in a detector.}) file and response file. This will allow us to perform X-ray and mid-IR simultaneous fit too. 
%%delete:Note that circumnuclear stellar component could be properly accounted for using a stellar library. However, its inclusion makes an impact on the resulting parameter estimates at mid-IR (Gonzal\'ez-Mart\'in  et al. submitted).
%Sugerencia dca: The IRS spectrum could have a high contribution of galaxy emission which implies consider a stellar library to model. The inclusion of this library makes an impact on the resulting parameter estimates at mid-IR.
%delete:In order to investigate it, we compared the \emph{Spitzer}/IRS spectrum with ground-based high spatial resolution images.
%For make sure that \emph{Spitzer} spectrum is not contain extra-nuclear components we compared this with high-resolution photometric data in Figure \ref{fig:mid-IRSpec}} %we compare the \emph{Spitzer} spectrum and these photometric data.}
%%%%%%%%%%%%%%%%%%%%%%%%%%%%%%%%%%%%%%%%%%%%%%%%%%%%%%%

\section{The mid-IR and X-ray models}

We give here a brief summary on the characteristics of the models used to fit X-ray and mid-IR spectra. Both wavelengths carry information on the torus-like structure that obscures the accretion disc for certain viewing angles. Both models are produced using radiative transfer codes including the physics required to account for mid-infrared and X-ray main continuum features. The mid-IR models include re-emission due to dust while X-ray models mainly include reflection in neutral gas.

    \subsection{X-ray model}\label{subsec:modelXray}
The bulk of the AGN emission is produced in the accretion disc and emitted at optical and ultraviolet (UV) wavelengths. A portion of this emission is reprocessed by a corona of a hot electrons plasma close to the accretion disc that scatters the energy in the X-ray bands due to inverse Compton \citep[][and references therein]{Netzer15, Ramos-Almeida17}. This comptonization produces one of the three main components of X-ray spectra known as the intrinsic continuum. It is modelled by a power law with a spectral index ($\Gamma$) typically around 1.8-2.3 \citep[e.g.][]{Yang15}. This feature dominates the spectral emission above 2 keV and it is a distinctive signature of the AGN emission. 
%The power-law emission is a function of the plasma temperature ($kT$), optical depth and the cutoff energy ($E_{cut}$) \citep{Haardt91, Marinucci15}.  \citep[$\rm{N_H > 10^{24} \, cm^{-2}}$][]{Zdziarski95,Magdziarz95}
Some part of this primary emission is absorbed by the torus or the broad line region and another is reprocessed by a distant material (e.g. the inner walls of the torus) and it gives place to the second most relevant component, named Compton hump with a maximum of its emission at $\rm{\sim}$30 keV \citep[][]{Ricci11}.
The reflection component depends on the shape of the reprocessing material, both its geometry and density \citep{Ghisellini94}. This structure could be the torus and depends mainly on the geometrical covering factor of the reprocessed material and its average $\rm{N_H}$. The third component is the FeK$\rm{\alpha}$ emission line, which origin is the reflection of X-ray photons.
%The FeK$\rm{\alpha}$ emission line is due to the reflection of X-ray photons in Compton thin material.
The origin of the narrow FeK$\rm{\alpha}$ line might also be associated to the torus, while the broad FeK$\rm{\alpha}$ line is thought to be originated in the inner parts of the accretion disc \citep{Fabian98, Laor91}. 
%\textbf{The best case for the relativistic broadening} is found in MCG-6-30-15, where a red wing of the FeK$\rm{\alpha}$ line is attributed to reflection in the accretion disc at only a few Schwarzchild radii from the black hole \citep{Vaughan04}. 
This analysis is based on the hypothesis that the reprocessor is the torus, which seems to be the case for the vast majority of the sources \citep{Matt91}.

The reflection component of AGN has been studied through different models \citep[e.g.][]{Murphy09,Ikeda09,Brightman11,Liu15}.
In this work, we used a new grid of X-ray spectral templates called borus02 model presented by \citet{Balokovic18}. % The borus02 templates contain only the spectral components from reprocessing in the gas torus.
These templates were based on BORUS, a radiative transfer code which assumes a toroidal geometry of neutral gas. To generate the borus02 templates the geometry was simplified as a smoother toroidal distribution of gas. %Before: the BORUS geometry was simplified as a smoothed distribution of individual clouds that make up a torus.
This geometry approximation is represented as a uniform-density sphere with two conical polar cutouts with the opening angle as a free parameter, such as the one employed by \citet{Brightman11} \citep[see also][for more details]{Balokovic18}.

The borus02 model allows us to explore the following parameters of the torus: 1) the average column density ($N_{H_{tor}}$), 2) the relative abundance of iron ($\rm{A_{FeK\alpha}}$), and 3) the angular size ($\rm{\theta_{tor}}$). Additionally, borus02 considered the incident emission in the torus as a power law with index $\Gamma$ multiplied by an exponential cutoff ($e^{(-E/E_{cut})}$). Finally an additional parameter controls the viewing angle of the torus relative to the observer ($\rm{\theta_{inc}}$). Fig.\ref{fig:mid-IR_models} (top-right corner with orange labels) shows the geometry and parameters associated to borus02.
%%%%%%%%%%%%%%%%%%%%%%%%%%%%%%%%%%%%%%%%%%%%%%%%%%%%%%%%%%%
    \subsection{Mid-IR model}\label{subsec:modelmid-IR}
    
\begin{figure*}[ht!]
    \centering
    \includegraphics[width=2.0\columnwidth]{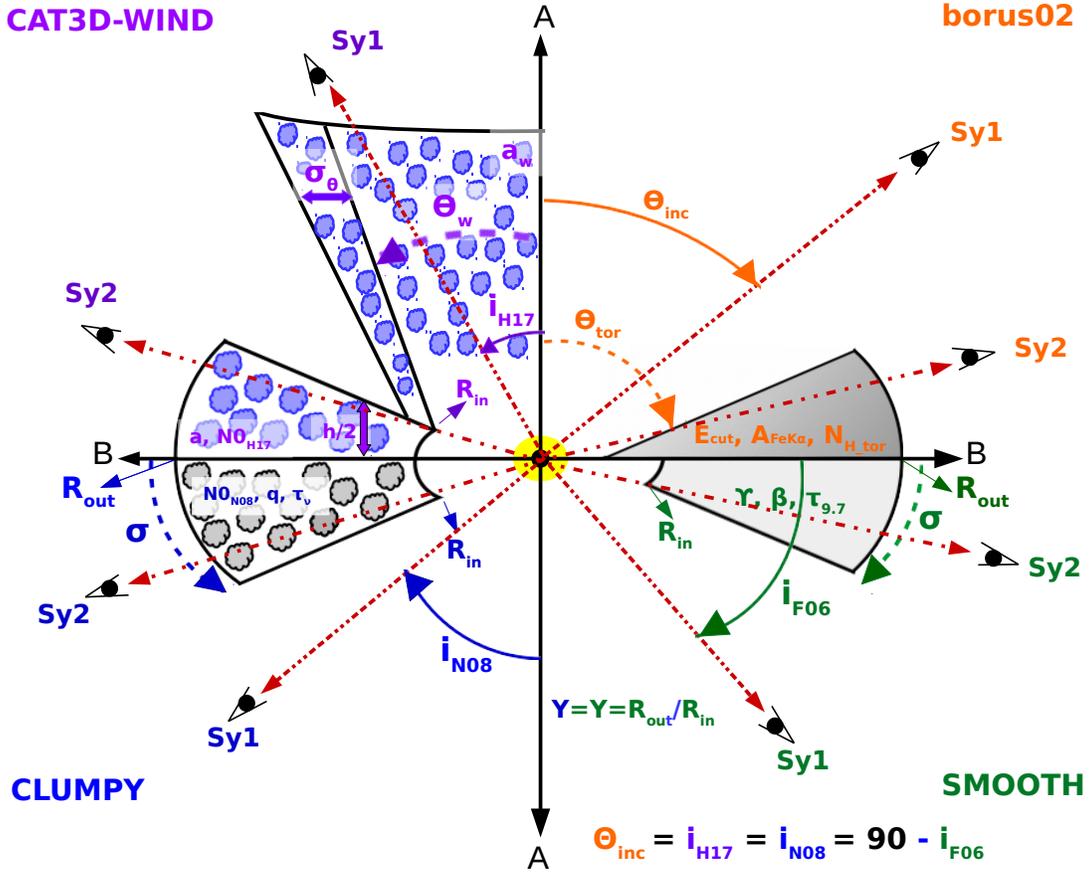}
    \caption{Geometry of the borus02 and the three mid-IR models used in this paper. The borus02 model by \citet{Balokovic18} is shown as a torus cut surface above the equatorial plane and filled in dark gray gradient. The smooth torus model by \citet{Fritz06} is shown as a torus cut surface below the equatorial plane and filled in light gray. The clumpy torus model by \citet{NenkovaB08} is shown as a torus cut surface below the equatorial plane filled with gray clouds. The CAT3D-WIND model by \citet{Hoenig17} is shown as a disc+wind cut surface above the equatorial plane and filled with blue clouds. Clouds represent models with clumpy distribution of dust. The parameters for each model are shown with different colors: borus02 (orange), Smooth (green), Clumpy (blue), CAT3D-WIND (purple), and borus02 (orange). In all of them we show an example of a view for a Sy1 and a Sy2.}%\footnote{Note that inner radius ($R_{in}$) is given as a function of the sublimation radius, i.e. $\rm{Y=R_{out}/R_{in}}$. Assuming a sublimation dust temperature of $\rm{T_{sub} = 1500\,K}$, The dust sublimation radius is $\rm{R_{in} = 1.3*(L_{AGN}/10^{46}\, erg s^{-1} )^{1/2}}$ for the smooth and the clumpy models and $\rm{R_{in} = 0.9*(L_{AGN}/10^{46}\, erg s^{-1})^{1/2}}$ \citep[ISM large grains][]{Hoenig10a} for the CAT3D-WIND model. See text for more details.}.}
    %The inclination angles of the Clumpy ($i_{N08}$), CAT3D-wind ($i_{H17}$), and borus02 ($\theta_{inc}$) models are measured clockwise. In the case of Smooth, the inclination angle ($i_{F06}$) is measured in the opposite direction. The half opening angle ($\sigma$) is measured equal for the Smooth and the Clumpy model. See text for more details.}}
    \label{fig:mid-IR_models}
\end{figure*}

The dusty torus has been the subject of several AGN models at mid-IR wavelengths that aimed to extract physical properties from SED. In this work, we used three of these SED grids to fit our mid-IR spectra: Smooth \citep{Fritz06,Feltre12}, Clumpy \citep{NenkovaA08,NenkovaB08}, and CAT3D-WIND \citep{Hoenig17}. These models are based on radiative transfer codes that use different geometrical distributions and compositions of dust. In Figure \ref{fig:mid-IR_models}, we show a cartoon that summarises the geometry assumed for these models. %The clumpy model is shown in the upper half torus with gray clouds and their parameters are shown with blue letters. The smooth model is shown in the lower half torus filled in gray and the parameters are shown with green letters. The two components (disc and wind) of the CAT3D-WIND model are shown with blue clouds and their parameters are shown with purple letter.
Below we give a short description for each model.

% and they are described as composition of power laws \citep{Granato94, Nenkova02}. Each SED produced by the CAT3D-WIND model has been broken down into three parameters: 1) the spectral index of the mid-IR emission, 2) the spectral index from near-IR to the mid-IR, and 3) the silicate feature optical depth. 

$\bullet$ \underline{Smooth model:} This model has a torus-like morphology. It was created by modelling a flared disc created as two concentric spheres, delimiting, respectively, the inner ($\rm{R_{in}}$) and the outer ($\rm{R_{out}}$) radius of the torus with the polar cones removed. It considers graphite grains with radius $\rm{a_G =  0.05\, \mu m}$ and sublimation dust temperature of 1500k to compute the $\rm{R_{in}}$ \citep[see Eq.5 from][]{Barvainis87}. It describes the dust density in polar coordinates \citep[see Eq.3 in][]{Fritz06} and allows to explore the following parameters of the torus: 1) the viewing angle of observer toward the torus ($\rm{i_{F06}}$), 2) the half opening angle ($\sigma$), 3) the exponent of the logarithmic elevation density distribution ($\gamma$), 4) the exponent of the power law of the radial profile of the density distribution ($\rm{\beta}$), 5) the equatorial %{\bf Donaji: estas segura que es ecuatorial? o promedio? Si es ecuatorial: Fritz06, Feltre12}
optical depth at 9.7 $\mu m$ ($\rm{\tau_{9.7}\, \mu m}$), and 6) the outer-to-inner radius ratio ($Y$).
%The SED produced by this model is defined in a $0.001 \mu m$ - $1000 \mu m$ wavelength range and it is describe as composition of power laws \citep{Granato94, Nenkova02}.

$\bullet$ \underline{Clumpy model:} The clumpy model considers a formalism where an AGN is surrounded by a toroidal distribution of dusty clouds. This assumes a standard Galactic composition (of $53\%$ silicates and $47\%$ graphite) of dust. Among them the most extensively used one is the Clumpy model of \citet{NenkovaB08} \citep[although see also][]{Hoenig10b} due to their large number of SEDs and probed ability to explain the mid-IR emission of low luminous \citep{Gonzalez-Martin17}, intermediate luminous \citep[e.g.][]{Ramos-Almeida09}, and high luminous \citep{Martinez-Paredes17} AGNs. The model parameters are: 1) the viewing angle ($i_{N08}$) respect to the polar plane, 2) the number of clouds in the equatorial plane of the torus ($\rm{N_{0_{N08}}}$), 3) the half angular width of the torus ($\sigma$), 4) the ratio between the inner and the outer radius ($\rm{Y=R_{out}/R_{in}}$), 5) the slope of the radial distribution of clouds described by a power law ($\rm{q}$), and the optical depth for the individual clouds ($\rm{\tau_{\nu}}$).

$\bullet$ \underline{CAT3D-WIND model:} The CAT3D-WIND model is built upon the hypothesis that the dust around the AGN consists of a geometrically thin disc of optically thick dust clumps and a outflowing wind described by a hollow cone composed by dusty clouds. The near-IR emission up to $\rm{\sim 5 \, \mu m}$ is due to an inflowing disc in the equatorial plane, while the main contributor to mid-IR emission is the polar dust. %This wind is launched in a puffed-up region of the inner hot part of the disc \citep[also see][]{Hoenig12}.
%In practice, this model contain two main components the clumpy disc and the polar dust.
%\citet{Hoenig10b} represented the clumpy disc with a torus-like geometry that decreases as the radius increases.
The distribution of the dust clouds in the disc is described with the following parameters: 1) the power law slope ($a$), 2) the inner radius ($\rm{R_{in}}$) that denotes the distance from the AGN in units of the sublimation radius, 3) the dimensionless scale height ($\rm{h}$) from the mid-plane of the disc of the vertical Gaussian distribution of clouds in units of the sublimation radius, and 4) the average number of dust clouds ($\rm{N_{0_{H17}}}$) along the equatorial LOS of the disc. The polar outflow is modeled as a hollow cone with the following parameters: 1) the radial distribution of dust clouds in the wind ($\rm{a_w}$), 2) the half-opening angle of the wind ($\rm{\theta_w}$), and 3) the angular width ($\rm{\sigma_{\theta}}$). Finally, two further parameters are common to both components, such as the inclination angle towards the observer ($i_{H17}$) and the ratio between the number of clouds along the cone and $\rm{N_{0_{H17}}}$ of the disc ($\rm{fwd}$). %The opacity of clouds is fixed to 50 and the outer radius of the polar distribution of clouds $\rm{R_{out}}$ is set to match the classical CAT3D value that does not result in artificial cutoffs in the mid-IR emission \citep{Hoenig10b}. 
This model considers a standard composition disc (similar to clumpy models) and an outflow composed by large grains.

The SEDs produced by the smooth and the clumpy models are defined in a $\rm{0.001 \, \mu m}$ - $\rm{1000\, \mu m}$ wavelength range and those produced by the CAT3D-WIND model cover a wavelength range between $\rm{0.01 \, \mu m}$ and $\rm{36,000 \, \mu m}$.

\subsection{Derived quantities}
    \subsubsection{Covering factor}
    \label{Sec:CF}
    
%The covering factor (hereafter \textit{Cf}) is the fraction of the sky that is covered by the obscuring material \citep{Ramos-Almeida17}. This parameter is capable of providing information about the physics of AGN feedback, the innermost regions around the SMBH, and the interaction of this with their host galaxy \citep[see further reference in][]{Balokovic18}. %Thus, it is worth to estimate this parameter and compare it for the different models. 

We calculate \textit{Cf} in both X-ray and mid-IR models. We use the relationship between the \textit{Cf} and $\theta_{tor}$ given by: \textit{Cf}$ = cos(\theta_{tor})$. Note that this is a simplistic approximation that assumes that the clouds take up most of the torus volume, following the prescription given by \citet{Balokovic18}. 

We can also calculate the \textit{Cf} using the mid-IR parameters. To derive the equation of the \textit{Cf} of Smooth model (\textit{Cf}$_{F06}$), we take into account the extinction coefficient, the density distribution along the radial and polar distances, the normalization constant, and $\beta$ value \citep[see equation 3 from][]{Fritz06}. Note that we assumed a $\beta = 0$ because this number only have two values in the SED provided (0 or 1) and we obtained a value close to zero for our best fit. For the Clumpy model we calculate this \textit{Cf} (\textit{Cf}$_{N08}$) using the equation 9 from \citep{NenkovaA08} and the angular distribution of clumps \citep{Feltre12}. A similar equation is used to calculate the \textit{Cf} for each component (wind and disc) for the CAT3D-wind model, with the total \textit{Cf} as the sum of the two components. Note that the \textit{Cf} of the wind, is calculated as the subtraction of two toroidal structures as in \citet{Hoenig10a} with a half opening angle of $\rm{\theta_{w} + \sigma_{\theta}}$ and $\rm{\theta_{w}}$, respectively. Using our notation for each parameter, these are the equations required to compute \textit{Cf}:

    \begin{itemize}
        \item Smooth model (case $\beta = 0$)
            \begin{equation}
             \textit{Cf}_{F06} = \frac{ln(\tau_{9.7})}{\gamma}
            \end{equation}
        \item Clumpy model
            \begin{equation}
                \textit{Cf}_{N08} = 1 - \int_0^{\pi/2} cos \beta e^{N_{0_{N08}} e^{-\beta^2/\sigma^2}} d\beta
            \end{equation}
        \item CAT3D wind model
            \begin{equation} \label{cfw}
            \begin{split}
                \textit{Cf}_{H17} = 1 + \int_0^{\pi/2} cos \theta e^{N_{0_{w}} e^{-\theta^2 / \theta^2_w }} d\theta  \\ - \int_0^{\pi/2} cos \theta e^{N_{0_{w}} e^{-\theta^2 / (\theta_w + \sigma_{\theta})^2 }} d\theta
                \\ - \int_0^{\pi/2} cos \theta e^{N_{0_{d}} e^{-\theta^2 / \sigma^2_d }} d\theta
            \end{split}
            \end{equation}
            \noindent where $\sigma_{d} = \arctan{h/2}$.
    \end{itemize}
    
    \subsubsection{Dust Mass}

We also estimate the total dust mass ($\rm{M_{tor}}$) using the parameters obtained for each model. This value is obtained from the integration of the density distribution of dust over the volume. We use the equation 9 from \citet{Mor09} given $q = 2$ for the Clumpy model. We follow the equations in table 1 from \citet{Hoenig10a} to compute the mass for the CAT3D disc-wind model. We sum up the contributions of the disc and the wind, with the latter obtained as the subtraction of two toroidal distribution with angular width of $\rm{\theta_{w} -  \sigma_{\theta}}$ and $\rm{\theta_{w}}$. Note that we analytically derive the mass equations assuming $\beta=0$ for the Smooth model and $q=2$ for the Clumpy model because these values are close to the results that we find for our object (see Table \ref{Table:Individual_fit} and Section \ref{sec:resultsfit}). Using our notation for each parameter, the equations to compute the total dust mass are:

    %Given the torus parameters, we can estimate the torus mass from 
    \begin{itemize}
        \item Smooth model ($\beta = 0$)
        \begin{equation}
            M_{tor} (F06) = \frac{4 \pi \tau_{9.7}}{3\kappa \gamma} \frac{(R^3_{out}-R^3_{in})(1-e^{-\gamma})}{R_{out}-R_{in}}
    \end{equation}
    where $\kappa$ is extinction coefficient in the Milky Way.
    \item Clumpy model ($q=2$)
    \begin{equation}
        M_{tor} (N_{0_{N08}}) = 4 \pi \sin (\sigma) N_{0_{N08}}*N R_{in}^2 Y \frac{Y}{2 log_{10} Y}
    \end{equation}
    where $N = N_H*A_{\nu}*mH$ is the $\rm{N_H}$ multiplied by the extinction due to dust (obtained from $\tau_{\nu}$ and assuming a constant dust-to-gas ratio) times the hydrogen mass in units of kg for a single cloud.
    
    \item CAT3D-wind model\footnote{Note that these equations are derived from \citet{Hoenig10a} assuming $b=1$ as in \citet{Hoenig17}.}
    \begin{equation}
        \begin{split}
         & M_{tor} (H17) =  \frac{N \sqrt{\pi}}{R^2_{cl;0}} \\ &\quad \times (N_{0_{w}} (f_{(\theta_{w} - \sigma_{\theta})} - f_{\theta_w}) + N_{0_{d}} f_{\sigma_{d}} )
         \end{split}
    \end{equation}
    where $f_{\theta_0}$ function is defined as \citep{Hoenig10a}
    \begin{equation}
        \begin{split}
         & f_{\theta_0} = \theta_0 e^{-\theta^2_0/4} \\ &\quad \times \left(Erf\frac{\pi - i \theta^2_0}{2 \theta_0} + Erf\frac{\pi + i \theta^2_0}{2 \theta_0} \right)
        \end{split}
    \end{equation}
    \end{itemize}

The constant dust-to-gas ratio relation assumed is $N_{H} = 1.9 \times 10^{21} * 1.086* \tau_{\nu}$ \citep{Bohlin78}. 

\section{Spectral fitting}\label{sec:fitting}

Spectral fitting is performed using the XSPEC fitting package. XSPEC is a command-driven, interactive, spectral-fitting tool within the HEASOFT\footnote{https://heasarc.gsfc.nasa.gov} software. XSPEC has been used to analyze X-ray data such as \emph{ROSAT}, \emph{ASCA}, \emph{Chandra}, \emph{XMM}-Newton, \emph{Suzaku}, \emph{NuSTAR}, or \emph{Hitomi}. XSPEC allows users to fit data with models constructed from single emission components coming from different mechanisms and/or physical regions. XSPEC already includes a large number of models but new ones can be incorporated using the {\sc atable} task. The borus02 templates have been included in XSPEC using this tool. In particular we use the $\rm{\chi^2}$ statistics (through the standard $\chi_{r}^2 = \chi^2/$d.o.f., where d.o.f is the number of degrees of freedom which is equal to the number of data bins in the spectrum minus the number of free parameters) and we assess the goodness-of-fit performing a test to reject the null hypothesis that the observed data are drawn from the model. % \textbf{(by including the reduced-$\chi^2 = \chi^2/$d.o.f., where d.o.f is the number of degrees of freedom and null hypothesis probability)}. 
The parameter confidence regions are found by surfaces of constant delta statistic from the best-fit value ({\sc error} task). Finally, XSPEC also allows us to find simultaneous confidence regions of multiple parameters to study the degeneracy among parameters. %Thus, XSPEC provides a wide range of tools to perform spectral fitting to data, being able to parallel processes in order to \textbf{speed them up}. 

X-ray data and models used in this analysis are already formatted to be used within XSPEC. To use these capabilities for the mid-IR spectra (and simultaneous X-ray and mid-IR fitting), we converted the data (see Section \ref{sec:Data}) and models (see Section \ref{sec:midXspec} below) to XSPEC format. 

%\begin{figure*}[ht]
%\begin{center}
%\includegraphics[width=0.9\columnwidth]{Figures/Figure_NuStar.png}
%\caption{NGC\,3516 fit spectra using Borus model. The green lines show the total fit, while the yellow lines show our model (using Borus02).}
%\label{fig.FitXraySpectrum}
%\end{center}
%\end{figure*}

    \subsection{Mid-IR models in XSPEC}\label{sec:midXspec}
    
We converted the mid-IR models SED libraries to multi-parametric models within the spectral fitting tool XSPEC as an additive table. The basic concept of a table model in XSPEC format is that the file contains a N-dimensional grid of model spectra with each point on the grid calculated for particular values of the N parameters in the model. XSPEC will interpolate on the grid to get the spectrum for the parameter values required at that point in the fit. To adapt mid-IR models we firstly created a one-parameter table (in fits format) associated to all the SEDs using the {\sc flx2tab} task within HEASOFT. Note that each of the SEDs have been interpolated using 5,000 steps between the minimum and maximum wavelengths due to the need of equally spaced SEDs. We then, wrote a python routine to change the headers associating each SED to a set of parameters. This model has the free parameters described in Section \ref{subsec:modelmid-IR} plus redshift and normalization.
For the clumpy model we were not able to obtain a XSPEC model using the entire SED library due to the unpractical size of the final model (over 100\,GB). Instead of $\rm{N_0=}$[1-15] and $\rm{\sigma=}$[15-70] in steps of 1 and 5, respectively, we slightly constrained the number of clouds and the angular width of the torus to the ranges $\rm{N_0=}$[1,3,5,7,9,11,13,15] and $\rm{\sigma=}$[15,25,35,45,55,65,70], respectively. This is in order to recover a more transferable model ($\sim$6\,GB).
%Note that we were not able to obtain a XSPEC model using the entire SED library due to unpractical size of the final model (over 100\,GB). Instead, we slightly \textbf{constrained} the number of clouds and the angular width to the torus to the ranges $\rm{N_0=}$[1,3,5,7,9,11,13,15] and $\rm{\sigma=}$[15,25,35,45,55,65,70], respectively, to recover a more transferable model ($\sim$6\,GB).
Note that this does not affect our results since XSPEC interpolates between models to find the best solution. 

%The mid-IR can be partially contaminated by stellar emission from the host. This is, indeed the case of NGC\,3516 where 10\% contribution of stellar component has been found by performing spectral decomposition to the data. In order to be able to include this component, we include...

\subsection{The total model in XSPEC}

    We fit the mid-IR and X-ray spectra of IC\,5063 following a command sequence in XSPEC:
    \begin{equation}
    \footnotesize
        \begin{split}
        & phabs * (atable \lbrace borus02 \rbrace + zdust*zphabs * cabs * cutoffpl) \\ & + zdust*atable \lbrace midIR_{model} \rbrace
        \end{split}
    \label{eq1}
    \end{equation}
\noindent where $phabs$ is the foreground galactic absorption\footnote{In the case of IC\,5063 this value is fix to $\rm{0.067 \times 10^{22} \, cm^{-2}}$, obtained by the {\sc nh} tool within Heasoft.}. The model borus02\footnote{We used the $borus02\_v170323c.fits$ file from http://www.astro.caltech.edu/~mislavb/download/} accounts for the reflection component. The $\rm{ zdust * zphabs * cabs}$ represents the line-of-sight absorption at the redshift of the source. Following the recipe provided by \citet{Balokovic18}, we linked the $\rm{N_H}$ component to the $zphabs$ to take into account for the total extinction along the line-of-sight, including the Compton scattering losses. We realised that these X-ray absorbers are not evaluated at energies below $10^{-4}$\,keV. Therefore, mid-IR and X-ray simultaneous fit requires that the X-ray intrinsic emission is properly absorbed below those energies. For that reason we introduced a \emph{zdust} component to neglect any artificial contribution of this component to mid-IR wavelengths. This model is also used to incorporate foreground extinction at mid-IR wavelengths.
%fit requires that the X-ray intrinsic emission needs to be properly absorbed below that energies.
We fixed the $\rm{Ecut}$ parameter to $\rm{300.0\,keV}$ because our X-ray spectra only cover a wavelength range between 3-100 keV. Also, we fixed the $\rm{A_{FeK\alpha}}$ parameters to the solar value. We varied these parameters on the final fit but they did not produce any statistical improvement. Finally, the $midIR_{model}$ is one of the three mid-IR models described in section \ref{subsec:modelmid-IR}.
%The variation range of the $\Gamma$ parameter is larger compared to ctorus and etorus and the angular size of the torus has a larger range than etorus. 

Note that the main advantage of the borus02 templates for our analysis is that it allows to constrain several parameters closely linked to the mid-IR models (see following sections).
    
\begin{table*}[ht]
\def\arraystretch{1.1}
%\caption{Models parameters to IC\,5063 }
\begin{center}
\begin{footnotesize}
\begin{tabular}{lcccccccc}
\hline \hline
 & bS1 baseline model  & & bC1 baseline model & & bW1 baseline model & & bS2 baseline model \\
\cline{2-2} \cline{4-4} \cline{6-6} \cline{8-8}
          & borus02 $+$ Smooth & & borus02 $+$ Clumpy & & borus02 $+$ CAT3D-WIND & & borus02 $+$ Smooth  \\
 Parameter  & $i_{F06} = 90. - \theta_{inc}$ & & $i_{N08} = \theta_{inc}$ & & $i_{H17} = \theta_{inc}$ & & $i_{F06} = 90. - \theta_{inc}$ \\
(1) & (2) & & (3) & & (4) & & (5) \\
\hline
$\Gamma$ & 1.72 $\pm_{0.06}^{0.07}$ & & 1.74 $\pm_{0.07}^{0.06}$ & & 1.70 $\pm_{0.08}^{0.06}$ & & 1.72 $\pm_{0.07}^{0.07}$ \\
$log(N_{H_{tor}})$ & 24.00 $\pm_{0.06}^{0.07}$ & & 23.98 $\pm_{0.07}^{0.02}$ & & 23.90 $\pm_{0.08}^{0.11}$ & & 23.99 $\pm_{0.07}^{0.08}$  \\
$log(N_{H_{los}})$ & 23.25 $\pm_{0.02}^{0.02}$ & & 23.25 $\pm_{0.03}^{0.01}$ & & 23.26 $\pm_{0.03}^{0.03}$ & & 23.25 $\pm_{0.02}^{0.03}$ \\
$\theta_{tor}$ & 60.0 $\pm_{4.5}^{2.6}$ & & 78.3 $\pm_{12.4}^{0.4}$ & & $< 21.9$ & & 56.0 ($90. - \sigma$) \\
$\theta_{inc}$ & 75.4 $\pm_{1.6}^{1.3}$ & & 87.1* & & 30.80 $\pm_{0.34}^{0.30}$ & & 75.3 $\pm_{1.5}^{1.5}$ \\
$\sigma$ & 34.1 $\pm_{1.9}^{0.8}$ & & 34.9 $\pm_{14.6}^{0.2}$ & & - & & 34.0 $\pm_{1.0}^{2.7}$ \\
$Y$ & 14.1 $\pm_{0.2}^{0.2}$ & & $>95.6$ & & - & & 14.1 $\pm_{0.2}^{0.2}$ \\
$\tau_{9.7}$ & $> 9.27$ & & - & & - & &  $> 9.1$ \\
$\tau_{\nu}$ & - & & 49.3 $\pm_{3.8}^{0.7}$ & & 50.0* & & -\\
$\beta$ & 0.0 & & - & & 1.0* & & 0.0 \\
$N_{0_{N08}}$ or  $N_{0_{w}}$ & - & & 3.67 $\pm_{0.20}^{0.03}$ & & $>7.46$ & & -\\
$\sigma_{\theta}$ & - & & - & & $< 7.2$ & & -\\
 $\theta_{w}$ & - & & - & & $> 44.6$ & & -\\
$\gamma$, $q$ or $aw$ & $> 5.7$ & & 2.13 $\pm_{0.02}^{0.03}$ & & $< -2.5$ & & $> 5.6$\\
$N_{0_{d}}$ & - & & - & & $>9.97$ & & -\\
$a$ & - & & - & & $ > -0.5$ & & - \\
$h$ & - & & - & & $> 0.5$ & & - \\
$fwd$ & - & & - & & $> 0.72$ & & -\\
$\chi^2 /$d.o.f. & $681/647$ & & $708/647$ & & $730/645$ & & $682/648$ \\
 \hline \hline
Derived parameters  &  & &  & &  & &  \\
%\multicolumn{5}{c}{Derived parameters} \\
 \hline
$R_{in}$ (pc) & 0.23* & & 0.23* & & 0.16* & & 0.23* \\
%$R_{in}$ ($R_{subl}$) & 2.54 (0.17\,pc)  & 2.54 (0.17\,pc) & 1.75 (0.16\,pc) \\
$R_{out}$ (pc) & 3.40 $\pm$ 0.05 & & $\rm{> 23.9}$ & & 450* & &  3.4 $\pm$ 0.05 \\
$Cf_{Xray}$ & 0.50 $\pm^{0.04}_{0.07}$ & & 0.20 $\pm^{0.21}_{0.01}$ & & $<$ 0.92 & & 0.56$\pm^{0.04}_{0.07}$ \\
$Cf_{midIR}$ & $>0.4$ & & 0.66 $\pm$ 0.01 & & $>$ 0.4 & & $>0.4$ \\
$M_{tor}$ ($\times\, 10^5\, M_{\odot}$) & $>0.06$  & & $>$ 30.3 & & $<0.2$** & & $>0.06$ \\
 \hline \hline
\end{tabular}
\caption{The best-fit physical parameters of the torus models for IC\,5063. The columns 2, 3, and 4 show the resulting parameters from fits assuming that inclination angles from mid-IR and X-ray models are linked. In column 5 shows the resulting parameters from fit assuming that inclination and half opening angles from smooth and borus02 models are linked.
The values marked with * are fixed parameters. **Total mass calculated as the sum of wind and disc masses ($0.03 \times 10^5\,M_{\odot}$ from wind and $0.16 \times 10^5\, M_{\odot}$ from disc). Note that the $Cf_{X-ray}$ is calculated as $cos(\theta_{tor})$, while the $Cf_{midIR}$ depends on several parameters according with the mid-IR model chosen (see Section \ref{Sec:CF}).
%Where $N0^1 = N0_{N08}$ and $N0^2 = fwd*N0_{H17}$ The sublimation ratio of this source is 0.067 pc. Rout considering a standard ISM.
%Best-fit physical unlinked parameters of the torus models to IC\,5063. 
}
\label{Table:Individual_fit}
\end{footnotesize}
\end{center}
\end{table*} % En las tablas poner todos los parametros, incluir flujos y luminosidades totales intrinsecas %El covering factor de Balokovic es 0.90 % El covering factor de smooth se debe calcular como sinSigma, clumpy con el programa de python. El covering factor de Balokovick es coseno.

\begin{figure*}[ht!]
    \centering
    \includegraphics[width=1.\columnwidth]{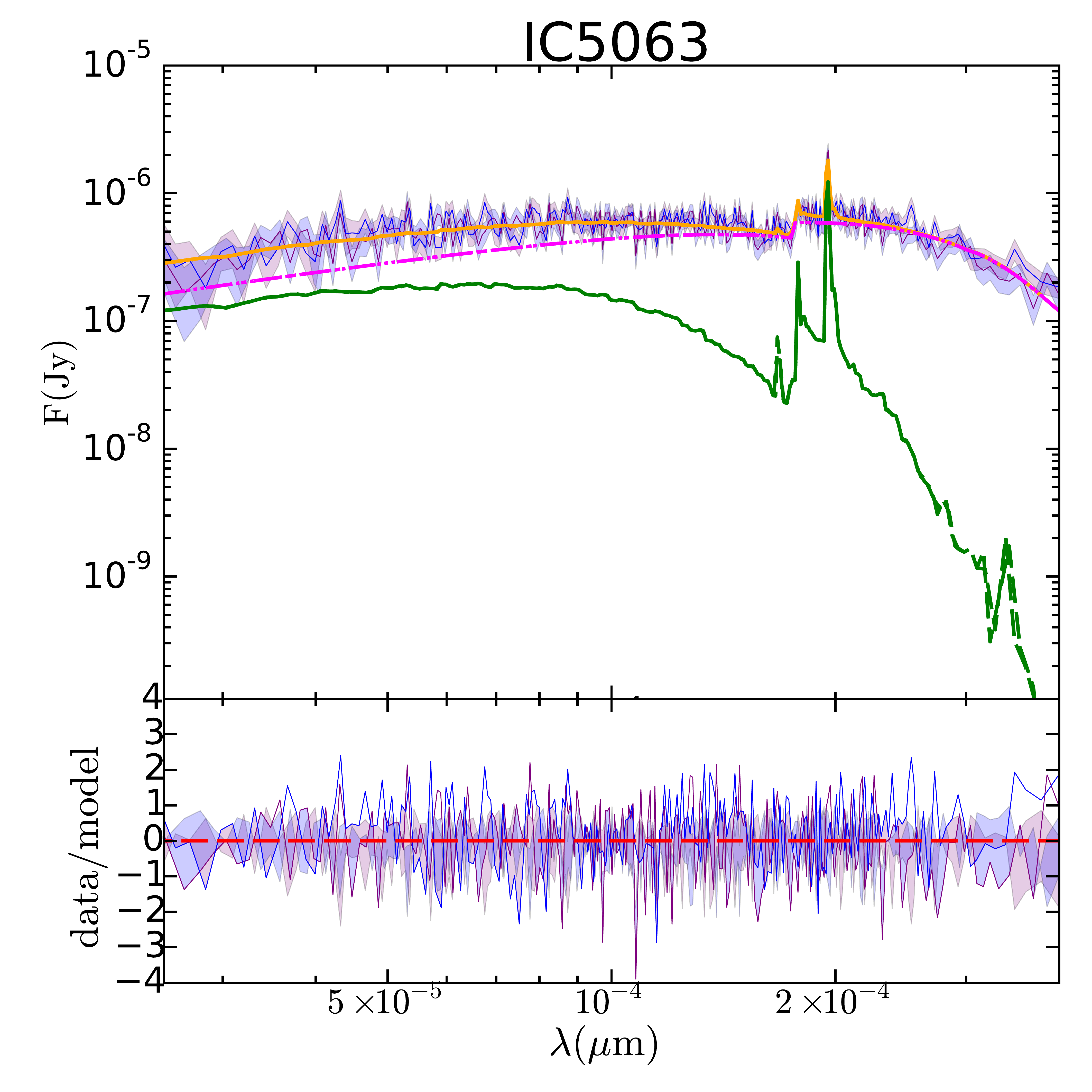}
    \includegraphics[width=1.\columnwidth]{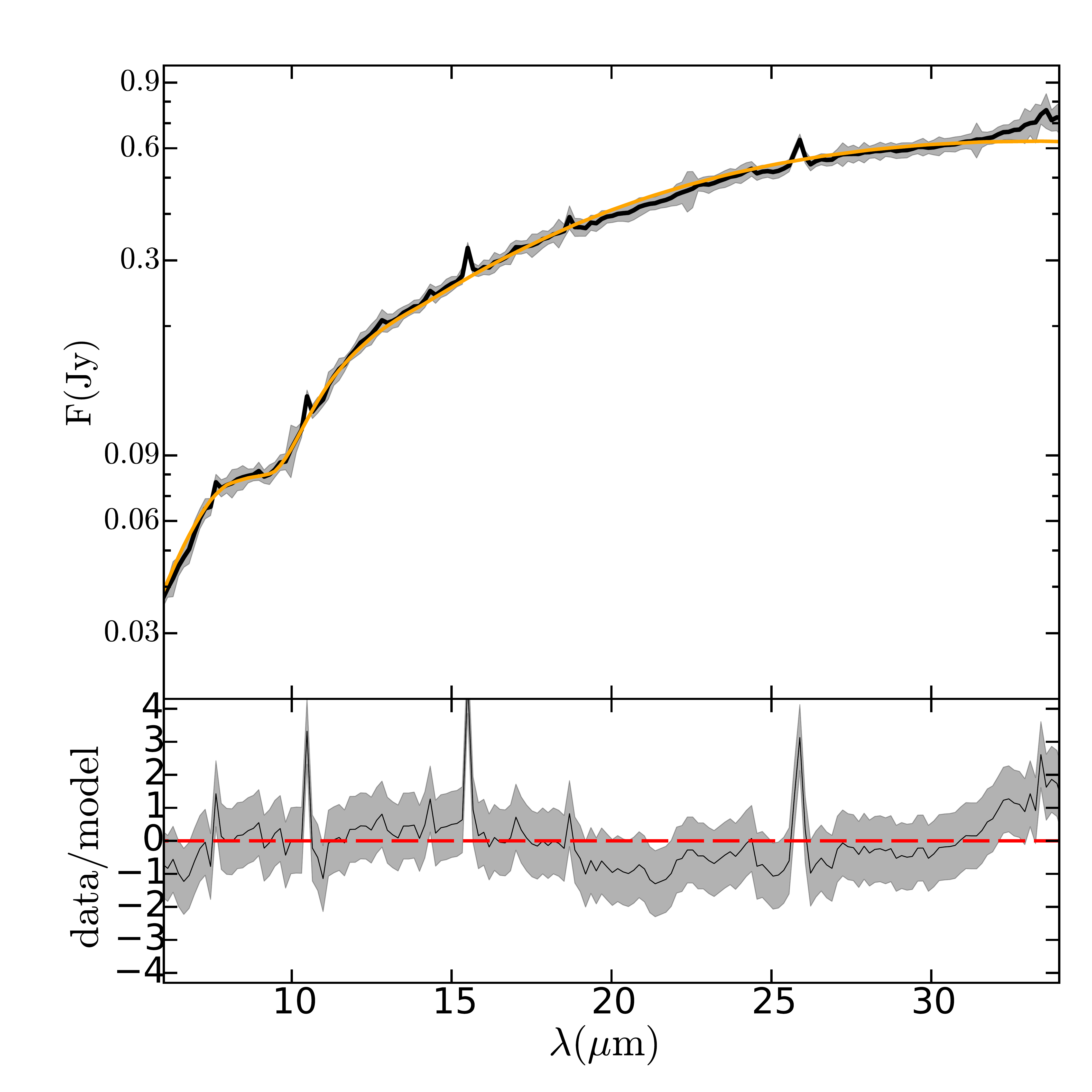}
    \caption{Unfolded spectra of IC\,5063. The orange solid lines are the best-fit obtained from the bS1 baseline model. Left: \emph{NuSTAR} spectra are displayed with blue and purple solid lines. The magenta and green dotted lines show the the absorbed power-law and the reprocessed components, respectively. Right: The \emph{Spitzer}/IRS spectrum is shown with a black solid line. The lower panels display the residuals between data and the best-fit model.}
    %Unfolded spectra of IC\,5063.The \emph{NuSTAR} (left) and \emph{Spitzer}/IRS (right) data are displayed with solid lines in blue and purple (left) and black (right), respectively. The overlaid solid lines are the best-fit obtained from the bS baseline model. Left figure:The absorbed power-law and reprocessed components (including fluorescent emission lines) are show with dotted lines. The lower panels display the residuals between data and the best-fit model.}
 \label{fig:bS_fit}
\end{figure*}

\begin{figure*}[ht]
    \centering
    \includegraphics[width=1.\columnwidth]{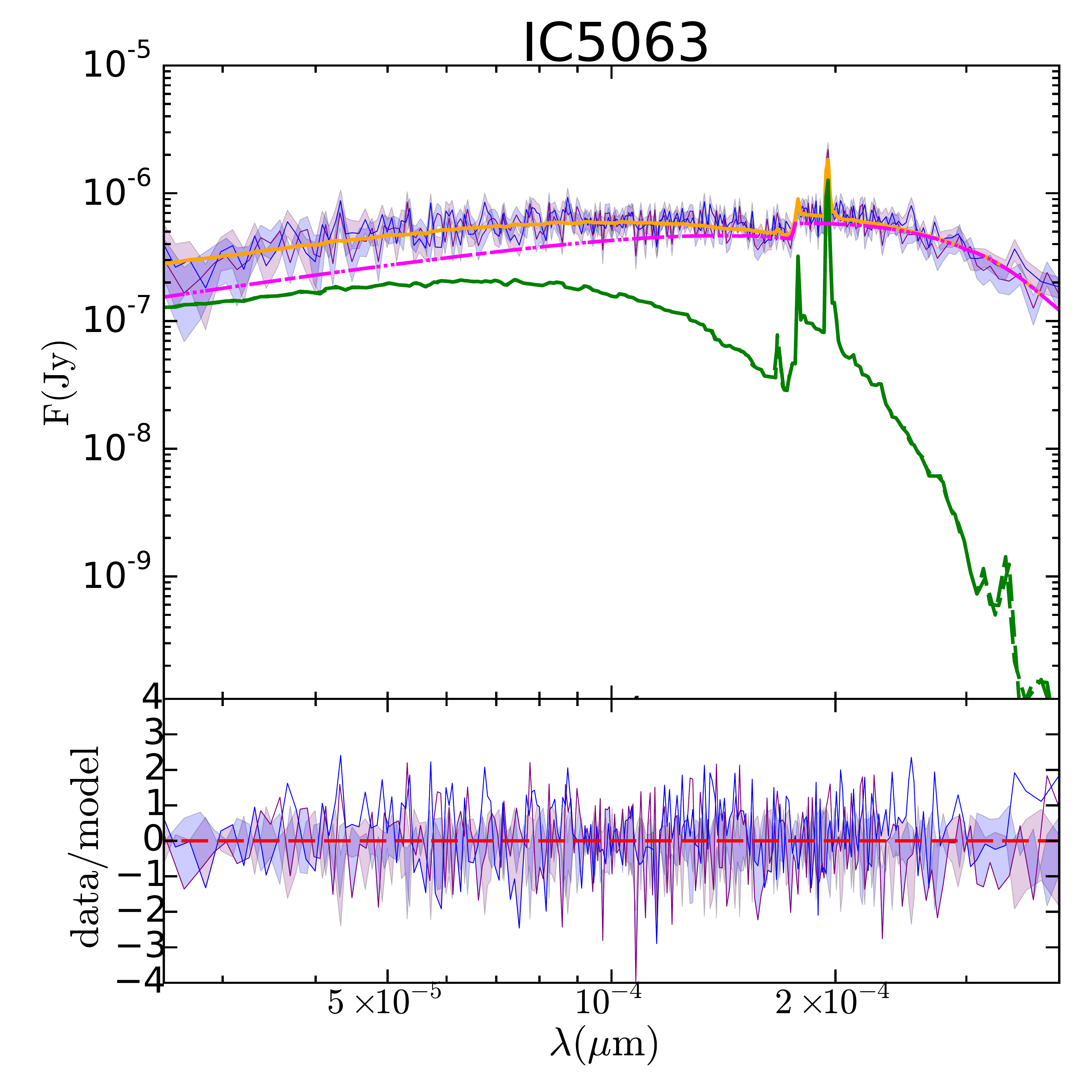}
    \includegraphics[width=1.\columnwidth]{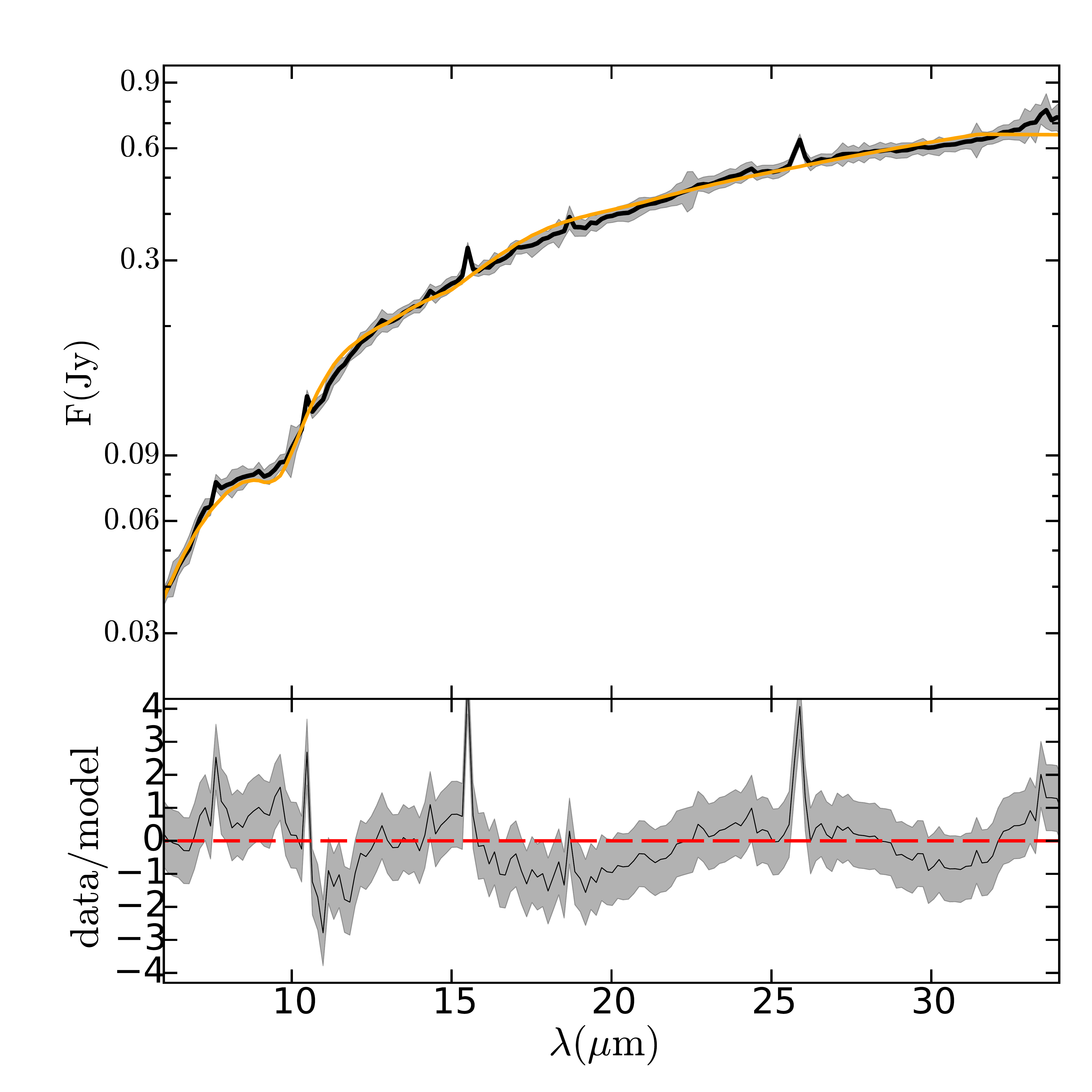}
    \caption{Same as Figure \ref{fig:bS_fit} but using the bN1 baseline model.}
    \label{fig:bSC_fit}
\end{figure*}

\begin{figure*}[ht]
    \centering
    \includegraphics[width=1.\columnwidth]{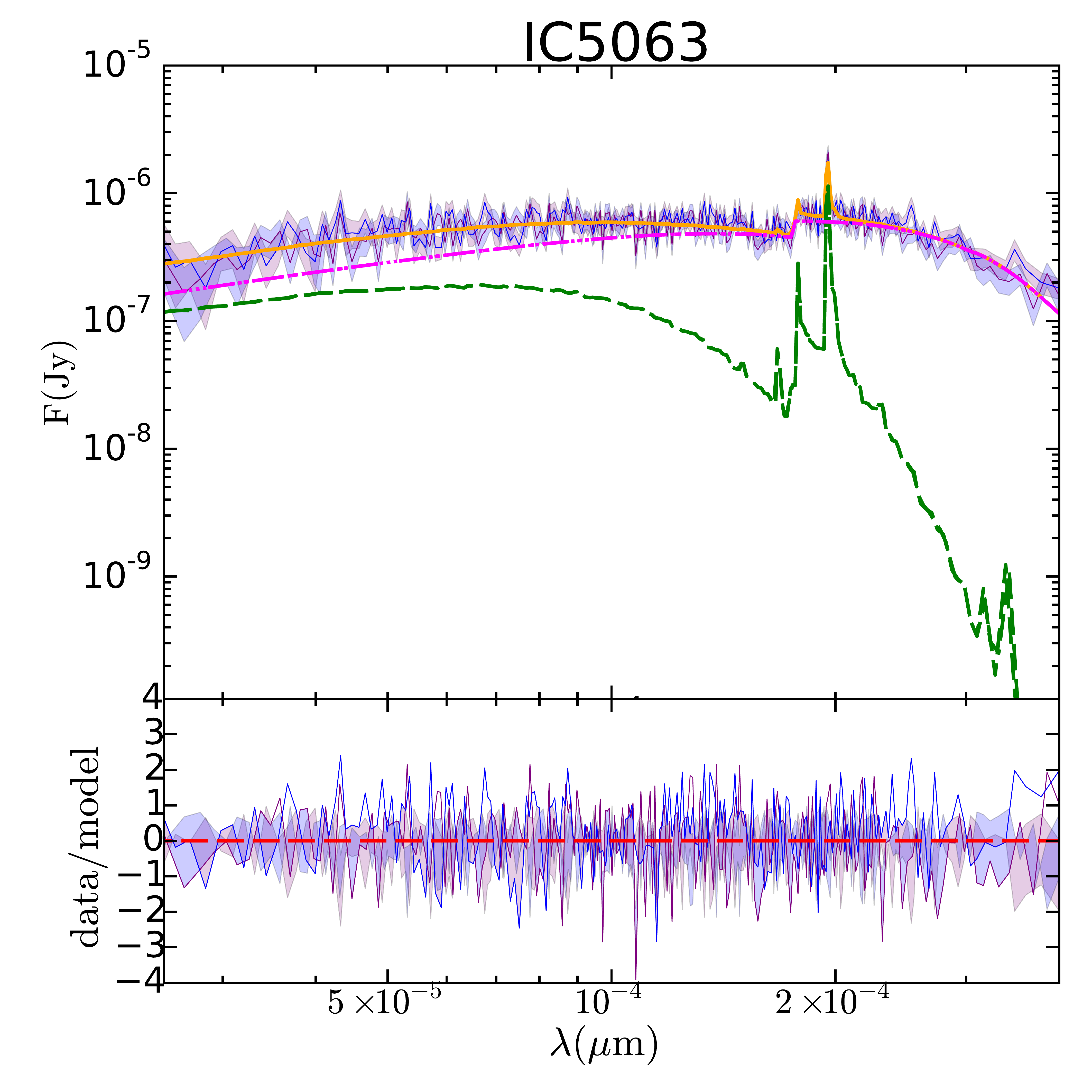}
    \includegraphics[width=1.\columnwidth]{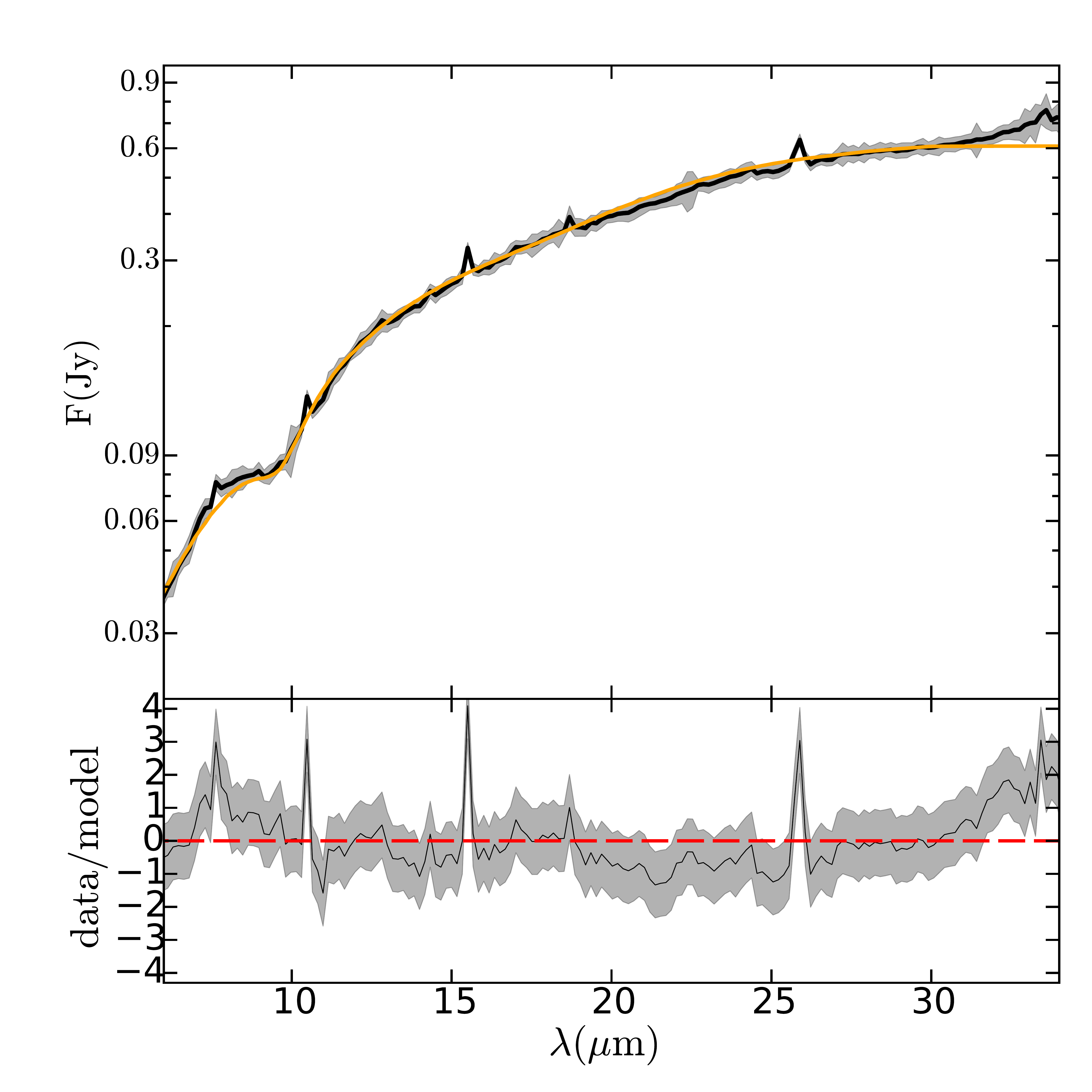}
    \caption{Same as Figure \ref{fig:bS_fit} but using the bW1 baseline model.}
    \label{fig:bSW_fit}
\end{figure*}

\section{Fitting results}\label{sec:resultsfit}
\subsection{Linking viewing angles}\label{sec:resultsfree}
    
%We firstly consider detached models expect for the viewing angle, that it is linked for both models to the same value and left free as a single parameter at both wavelengths. Therefore, in the subsequent analysis the viewing angle in mid-IR models ($i$) is linked to the viewing angle for borus02 model ($\rm{\theta_{inc}}$). Note that we used the convention of viewing angles as shown in Fig. \ref{fig:mid-IR_models} (i.e. $i_{F06} = 90. - \theta_{inc}$,  $i_{N08} = \theta_{inc}$, and $i_{H17} = \theta_{inc}$).Table \ref{Table:Individual_fit} shows the resulting values for each parameter after fitting simultaneously the \emph{NuSTAR} spectra using the borus02 model and \emph{Spitzer} spectrum with each of the three mid-IR models (smooth torus, clumpy torus, and wind-disc models at columns 2, 3, and 4, respectively). Hereafter, we refer to these combinations of the borus02 model with either the Smooth, Clumpy, or CAT3D-wind model as bS, bC, and bW baseline models, respectively.
We firstly consider that the only parameter linked between mid-IR and X-ray models is the viewing angle. Therefore, we linked to the same value the mid-IR and X-ray viewing angles as shown in Fig. \ref{fig:mid-IR_models} (i.e. $i_{F06} = 90 - \theta_{inc}$,  $i_{N08} = \theta_{inc}$, and $i_{H17} = \theta_{inc}$).
%The viewing angle in mid-IR models ($i_{model}$) is linked to the viewing angle of the borus02 model ($\rm{\theta_{inc}}$) as shown in Fig. \ref{fig:mid-IR_models} (i.e. $i_{F06} = 90 - \theta_{inc}$,  $i_{N08} = \theta_{inc}$, and $i_{H17} = \theta_{inc}$).Then the $\rm{\theta_{inc}}$ will be considered as left free parameter. 
Table \ref{Table:Individual_fit} shows the resulting values for each parameter after fitting simultaneously the \emph{NuSTAR} spectra using the borus02 model and the \emph{Spitzer}/IRS spectrum with each of the three mid-IR models (smooth torus, clumpy torus, or clumpy wind-disc models at columns 2, 3, and 4, respectively). Hereafter, we refer to these combinations of the borus02 model with either the Smooth, Clumpy, or CAT3D-wind model as bS1, bC1, and bW1 baseline models, respectively.
% (see Cols. 2, 3, and 4 in table). %Note that the rest of this subsection is dedicated to discuss the results of these combined models.

%In table \ref{Table:Individual_fit}, we show the resulting value of each parameter after fitting the SED using the borus02 model with each of the three mid-IR models. We consider that inclination angles in both wavelengths must be related. Therefore, we linked the inclination angles of each mid-IR model to the value in the X-ray model. We used the aperture angle direction observed in Fig. \ref{fig:mid-IR_models} %we used the aperture angle direction reported in previous works
%(i.e. $i_{F06} = 90. - \theta_{inc}$,  $i_{N08} = \theta_{inc}$, and $i_{H17} = \theta_{inc}$).

Regarding the X-ray parameters, we found that: (1) the $\rm{logN_{H_{tor}}}$ is independent from the mid-IR model selected; (2) the $\rm{\Gamma}$ shows slight changes depending on the model; and (3) the $\rm{\theta_{tor}}$ and $\rm{\theta_{inc}}$ strongly depend on the mid-IR model used. Note that the viewing angle $\rm{\theta_{inc}}$ is constrained when using the bS1 and bW1 baseline models, and both angles are consistent with a Sy2.
%The latter proves the importance of selecting the best model for the mid-IR data to derive meaningful quantities.
%the average column density of the torus from X-rays, the spectral index of the incident radiation, the angular width of the torus and the viewing angle toward the observer 
   
The bS1 baseline model has four free mid-IR parameters; two being constrained, other two ($\rm{\gamma}$ and $\rm{\tau_{9.7}}$) are close to the upper limit defined by the model; and $\rm{\beta}$ is set to $\rm{\beta=0}$, giving better results than $\rm{\beta=1}$. Only $Y$ is close to the upper limit among the five free parameters for bC1 baseline model, being the other free parameters well constrained. Finally, among the seven free parameter of the bW1 baseline model, five are upper limits and two are lower limits. 

While a direct comparison between mid-IR parameters from each model is indeed challenging, we compare some mid-IR parameters among the bS1, bC1, or/and bW1 baseline models, such as the $\rm{\sigma}$, $Y$, $N_{0_{N08}}\,(N_{0_w})$ and the power law indices of the dust radial distributions ($q$, $\gamma$ or $aw$). In particular, we can compare the bC1 and the bS1 baseline model in terms of $\rm{\sigma}$ and $\rm{Y}$ parameters. Similar results are obtained for $\rm{\sigma}$ with both models. The bC1 baseline model shows a large value for the $\rm{Y}$ parameter that implies\footnote{Note that the inner radius of both models is set to the same value.} a torus size $\rm{> 24\,pc}$, compared to $\rm{\sim 3.4\,pc}$ for the bS1 baseline model. The latter is in better agreement with more recent works \citep[see references in][]{Ramos-Almeida17}. We also computed the number of clouds along the wind using $fwd$ and $N_{0_d}$ parameters ($N_{0_w} = fwd*N_{0_d}$) for the bW1 baseline model. This number is larger than the number of clouds in the equatorial LOS for the torus obtained for the bC1 baseline model (i.e. $N_{0_{N08}}$). %The three models have a parameter describing the slope of the radial dust distribution ($\rm{\gamma}$, $\rm{q}$, and $\rm{a_w}$). 
We found that $\rm{a_w}$ (bW1) is very similar to $\rm{q}$ (bC1), while $\rm{\gamma}$ has a higher value.
   
We used the reduced $\rm{\chi^2}$ statistic value to assess the goodness-of-fit for each model.%\footnote{XSPEC performs a test to reject the null hypothesis that the observed data are drawn from the model. After the fit, XSPEC writes out the reduced $\rm{\chi^2}$.}. 
The $\rm{\chi^2/d.o.f.}$ for the bS1, bC1, and bW1 baseline models are reported in table \ref{Table:Individual_fit}.
Note that there are not large differences between $\rm{\chi^2/d.o.f.}$ from these three baseline models, although the bW1 baseline model shows a larger $\rm{\chi^2/d.o.f }$ than the other two baseline models and a slightly better $\rm{\chi^2/d.o.f }$ is obtained with the combination of the borus02 and smooth models (bS1 baseline model). In Figure \ref{fig:bS_fit}, we show the IC\,5063 spectra and the resulting fit using bS1 baseline model. Note that the bS1 baseline model better reproduces the [7-10] $\mu m$ wavelength range compared with bC1 and bW1 baseline models (Figures \ref{fig:bSC_fit} and \ref{fig:bSW_fit}).
%%%%Note that the larger $\rm{\chi^2/d.o.f.}$ is obtained for the bW baseline model, while the best $\rm{\chi^2/d.o.f.}$ is obtained with the bS baseline model. Therefore, the combination of the borus02 and smooth models (bS baseline model) is the best choice to fit the mid-IR and Xspec spectra of IC\,5063 (see Figure \ref{fig:bS_fit}). Particularly, the bS baseline model fits the [7-10] $\mu m$ range compared with bC and bW baseline models (see also Figures \ref{fig:bSC_fit} and \ref{fig:bSW_fit} in Appendix \ref{sec:Appendix2}.)

%\textbf{In Figure \ref{fig:bS_fit} we show the unfolded \emph{NuSTAR} and \emph{Spitzer} spectra and the best-fit using the bS baseline model. Note that the [7-10] $\mu m$ range is best fit using this option. }
%\textbf{We found that the bS baseline model fitting best the [7-10] $\mu m$ range than bC and bW baseline models (see Figures \ref{fig:bS_fit} (right), and figures \ref{fig:bSC_fit} and \ref{fig:bSW_fit} in Appendix). 
%In Figure \ref{fig:bS_fit} we show the unfolded \emph{NuSTAR} and \emph{Spitzer} spectra and  } {\bf NOTA OGM: Quizas aqui nos falta mostrar los fits y decir en que rango espectrales es mejor Fritz respecto a los otros?}
     
We also explored the cases in which the direction of the inclination angle for the mid-IR models can be inversely related to the inclination angle for the X-ray band (i.e. $i_{F06} = \theta_{inc}$,  $i_{N08} = 90 - \theta_{inc}$, and $i_{H17} = 90 - \theta_{inc}$). This scenario will represent a reflector neutral gas that fills up the gaps where the mid-IR emitting dust is not present. We also obtained the statistic values for these cases and we compared them with those obtained above (i.e. direct link between viewing angles reported in table \ref{Table:Individual_fit}). We found that this interpretation of the viewing angle is worse than that assumed before for the bS and bC1 baseline models, obtaining a $\chi^2/d.o.f.$ of 685$/$647 and 754/648 (i.e. $\Delta \chi^2/d.o.f.$ 4 and 6 compared to those reported in the table), respectively.
Interestingly, in the case of the bW1 baseline model, we found that this new link between viewing angles is an improvement with $\chi^2/d.o.f.$ of 718/645 (i.e. $\Delta \chi^2/d.o.f. = -12$). %Therefore, under the disc-wind model assumption, the favoured scenario suggests that the gas producing the reflection at X-rays and the dust emitting at mid-IR are decoupled.

   % \subsubsection{Covering factor and Torus Mass}
%\textit{Cf} (X-ray), \textit{Cf} (mid-IR)
We reported the covering factors \textit{Cfs} obtained from X-ray models using bS1, bC1, and bW1 baseline models in table \ref{Table:Individual_fit} (quoted as $Cf_{Xray}$). Both bS1 and bC1 baseline models give consistent \textit{Cfs} within error bars while the bW1 baseline model gives a higher \textit{Cf}. Table \ref{Table:Individual_fit} also reports the \textit{Cfs} obtained using the mid-IR parameters (denoted as $Cf_{midIR}$). \textit{Cf} for different baseline models are consistent with each other. A comparison among $Cf_{Xray}$ and $Cf_{midIR}$ shows compatible results for the bS and the bW1 baseline models. However, the \textit{Cf} obtained from X-rays is larger than that obtained from mid-IR for the bC baseline model.

\begin{figure*}[ht!]
    \centering
    \includegraphics[width=2.1\columnwidth]{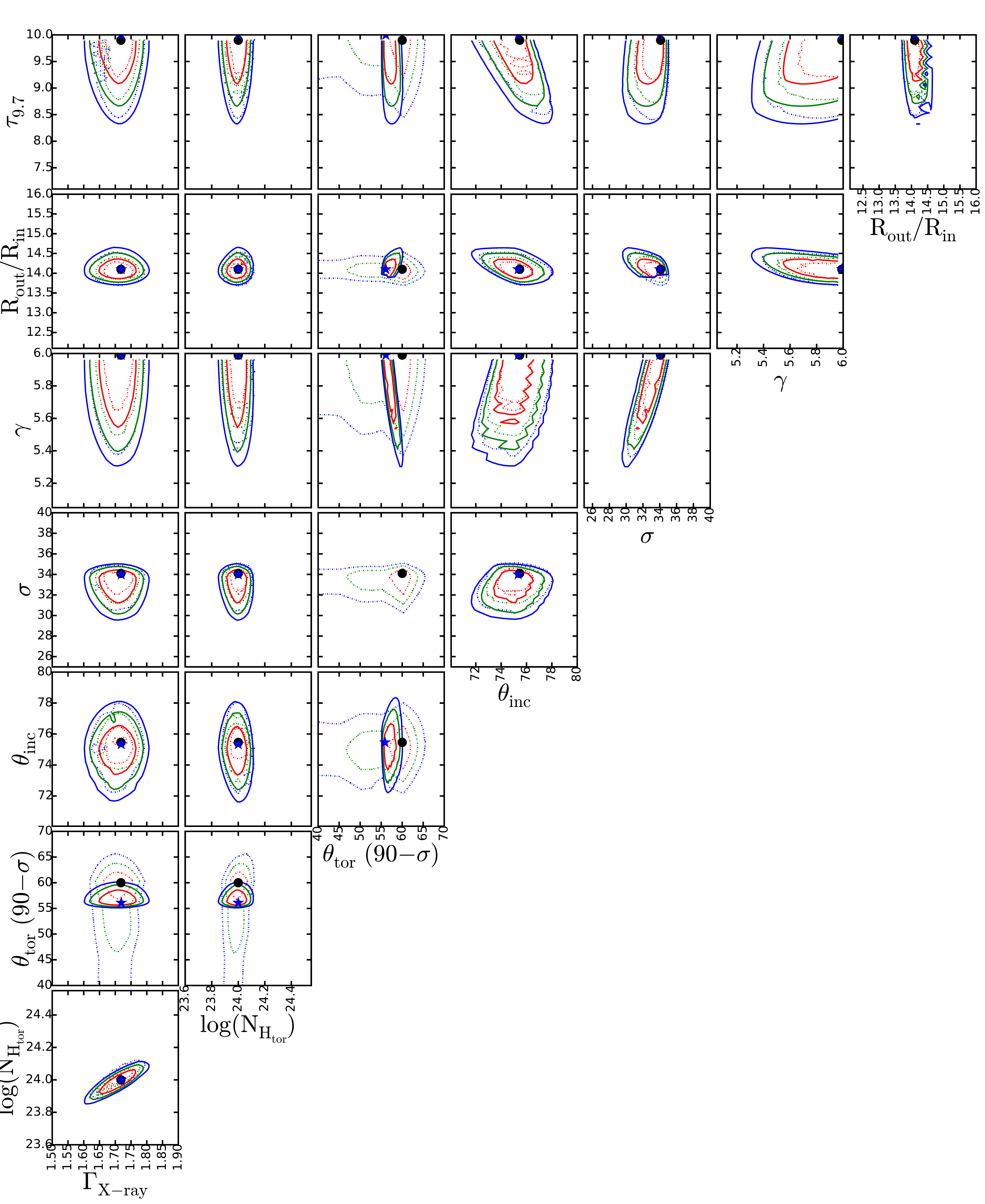}
    \caption{Two-dimensional $\Delta_{\chi^2}$ contours for the resulting free parameters when we used the bS1 (dotted lines) and the bS2 (solid lines) baseline models to fit IC\,5063. The red, green and blue (dotted and solid) lines show the contours at 1$\sigma$, 2$\sigma$, and 3$\sigma$, respectively. The black circles and blue stars are the resulting values for each parameter using the bS1 and bS2 baseline models, respectively. Notice that these values are reported in Columns 2 and 4 of Table \ref{Table:Individual_fit}.
    %Dimensional $\Delta_{\chi^2}$ contours for free parameters for IC\,5063. The red, green and blue lines are 1$\sigma$, 2$\sigma$, and 3$\sigma$. The blue stars and black circles show the best-fit values before and after linking the half opening angle of the torus (reported in Table \ref{Table:Individual_fit}).
    }
    \label{fig:ContoursFritzfree}
\end{figure*}

Finally, we check for the degeneracy among parameters of the fit. For this purpose we used the best fit baseline model obtained (i.e. bS). Figure \ref{fig:ContoursFritzfree} shows the two-dimensional $\rm{\chi^2}$ distribution for each free parameter (dotted lines). We found that most parameters are well constrained within the $\rm{3\, \sigma}$ contours. The most controversial parameter is $\theta_{tor}$ which we cannot yet restrict, being in the range [10-70] at the 2$\sigma$ level. However, note that $\sigma$ parameter from the Smooth model is constrained and both parameters ($\sigma$ and $\theta_{tor}$) could be directly linked (see below).
 %   \textbf{Aqui pongo que podemos ligar sigma y thetator, como se comparan de acuerdo con mi tabla.}
    
%Está figura esta hecha con el programa: PlottingSteppar_table2.py que está en: /home/donaji/Dropbox/XRAY_DATA/BALOKOVICDATA/IC5063

    \subsection{Linking both viewing and half opening angles} \label{sec:resultslink}
    
 %According to our previous results, the combination of borus02 and Smooth models (bS baseline model) is the best fit to the mid-IR and X-ray spectra of IC\,5063. Yet, this fit does not restrict the half opening angle of the torus inferred from X-ray, (i.e., the $\theta_{tor}$). However, the half opening angle inferred from the mid-IR model, i.e., $\rm{\sigma}$ is \textbf{constrained}. 
 
 We test here if we can link both the viewing angle and the half opening angle in the bS1 baseline model. Conforming to the definitions of the opening angles in the smooth and the borus02 models, the link between both parameters is $\rm{\theta_{tor} = 90 - \sigma}$. Hereafter, we refer to this new combination as bS2 baseline model.
 % According to the definitions of opening angles in X-ray and mid-IR models, the half opening viewing angle at X-rays corresponds to 90 degrees the half opening angle measured at mid-IR, i.e., $\rm{\theta_{tor} = 90 - \sigma}$. 
 In column 5 from Table \ref{Table:Individual_fit} we report the values obtained for bS2 baseline model.

These new parameters are consistent within the errors to those measured for the bS1 baseline model (reported in Table \ref{Table:Individual_fit}). This is the case even for $\sigma$ and $\theta_{tor}$. Note that the derived parameters \textit{Cf}\,(X-ray), \textit{Cf}\,(mid-IR), and \textit{$M_{tor}$} also remain the same compared to bS1 baseline model.
We compared $\chi^2/d.o.f.$ from bS and bS2 baseline models through the f-test, obtaining a probability 0.37 which is greater than $10^{-4}$. Therefore, we discard the hypothesis that a most complex baseline model (bS) is better to fit the spectra; i.e., the simpler baseline model (bS2) is enough to reproduce the data.
%{\bf OGM: Te falta comparar la chi2/dof con f-test para demostrar que este modelo más simple es suficiente. Es decir que añadir la posibilidad de que esten desligados no mejora el resultado. }
    
We also check the degeneracy among the parameters for bS2 baseline model. In Figure \ref{fig:ContoursFritzfree}, we show the two-dimensional $\chi^2$ distribution for each combination of parameters when use the bS2 baseline model (solid lines). Note that all parameters are constrained within the $\rm{3\,\sigma}$ contours. The advantage of linking them is that now the half opening angle is constrained for both fits. This slightly improves the calculus of the degeneracy of the parameters, showing smoother contours in Fig. \ref{fig:ContoursFritzfree} with no significant spoilage of the parameter restriction.
    
Other parameters that can be associated are $log (N_{H_{tor}})$ and $\rm{\tau_{\nu}}$ because both are associated to the density of the medium. However, the relationship between both parameters is not simple. We explore this possibility in the next section.
    %Note that $\rm{\theta_{tor}}$ and $\rm{\theta_{inc}}$ are directly linked to mid-IR models. Furthermore, $log (NH_{tor})$ can be associated to the opacity of the clouds $\rm{\tau_{\nu}}$ used within mid-IR Clumpy models.
    
    %\begin{table}[ht]
    %\def\arraystretch{1.1}
    %\caption{Models parameters to IC\,5063 }
    %\begin{center}
    %\begin{footnotesize}
    %\begin{tabular}{lc}
     %   \hline \hline
      %  Parameter & bS2 model \\
       %   & borus02 $+$ Smooth \\
          %& $i_{F06} = 90. - \theta_{inc}$ \\
          %& $\theta_{tor} = 90. - \sigma$  \\
        %  (1) & (2) \\
         % \hline
         % $\Gamma$ & 1.72 $\pm_{0.07}^{0.07}$  \\
         % $logNH_{tor}$ & 23.99 $\pm_{0.07}^{0.08}$  \\
        %  $logNH_{los}$ & 23.25 $\pm_{0.02}^{0.03}$  \\
        %  $\theta_{tor}$ & 55.9 ($90. - \sigma$) \\
        %  $\theta_{inc}$ & 75.3 $\pm_{1.5}^{1.5}$ \\
        %  $\sigma$ & 34.0 $\pm_{1.0}^{2.7}$ \\
        %  $Y$ & 14.1 $\pm_{0.2}^{0.2}$  \\
        %  $\tau_{9.7}$ & $> 9.1$ \\
        %  $\beta$ & 0.0  \\
        %  $\gamma$ & $> 5.6$ \\
        %  $\chi^2 /$d.o.f. & $682/648$ \\
        %  \hline \hline
        %  Derived parameters \\
        %  \hline
        %  $R_{in}$ (pc) & 0.23*  \\
        %  $R_{out}$ (pc) & 3.4 \\
        %  $Cf$ (X-ray) & 0.56 \\
        %  $Cf$ (mid-IR) & $>0.4$ \\
        %  $M_{tor}$ $(\times\, 10^5\, M_{\odot})$ & $>0.06$ \\
        %  \hline \hline
        %  \end{tabular}
        %  \caption{Best-fit physical linked parameters of the torus models to IC\,5063. The values marked with * are fixed parameters.}
%    \label{Table:Individual_fit_2}
%    \end{footnotesize}
 %   \end{center}
 %   \end{table}

\section{Discussion}\label{sec:Discussion}
In this work, we investigate the properties of the dusty torus of IC\,5063, exploring the combination of mid-IR and X-ray spectral fits. We discuss here if the same structure producing the mid-IR continuum can also describe the reprocessed emission at X-ray wavelengths (Section\,\ref{sec:discus1}), what are the resulting torus properties (Section\,\ref{sec:discus2}), and if the combination can better constrain the physical parameters of the dusty structure when they are used to fit simultaneous the \emph{Spitzer}/IRS and \emph{NuSTAR} spectra (Section\,\ref{sec:discus3}).

\subsection{Link between mid-IR continuum and reprocessed emission at X-ray wavelengths}\label{sec:discus1}

%Our first step to combine the information at both wavelengths was assuming that their model viewing angles are the same.
Our first step to combine the information at both wavelength ranges was to assume that the viewing angle of the torus is the same. For the bS1 and the bC1 baseline models, we found that the best options to link the mid-IR and X-ray viewing angles are $i_{F06} = 90. - \theta_{inc}$ and $i_{N08} = \theta_{inc}$, respectively. These options imply a scenario where dust and gas are in the same location (distributed along the equatorial plane). In the case of bW1 baseline model, we found that the best option is $i_{H17} = 90. - \theta_{inc}$. This scenario implies that most of mid-IR emission is in the equatorial plane.
%This scenario contradicts the built hypothesis of CAT3D-wind model because implies that most of mid-IR emission is in the equatorial plane.
Therefore, these three baseline models are consistent with the idea that most of the dust producing the mid-IR continuum emission is distributed in the equatorial plane where the torus has historically been located. Furthermore, the X-ray reflector under this scenario is also in the equatorial plane.

We reviewed and compared the reduced $\chi^2$ statistic values for each of the baseline models. From this analysis, we concluded that the best statistic is obtained when using the combination of borus02 (X-ray) and Smooth (mid-IR) models (the so-called bS1 baseline model) to fit the spectra at both wavelengths. Even though this baseline model has the best reduced $\chi^2$ it is not capable of restricting the half opening angle from X-ray ($\theta_{tor}$), but it can restrict the mid-IR half opening angle ($\sigma$). This issue is solved when both viewing and opening angles are linked (bS2 baseline model). We found that all the parameters can be constrained in bS2 baseline model using $\rm{\theta_{tor} = 90 - \sigma}$. The link between half opening angles suggests a common origin for both emissions. Indeed, the statistic does not improve if these two parameters are allowed to vary individually. Therefore, the bS2 baseline model, where the viewing and half-opening angles are tied together, is enough to explain the observations at both wavelengths.
The fact that the inclination and half-opening angles from mid-IR and X-ray are directly linked to the same value is consistent with previous results \citep{Farrah16}.
%%\textbf{The latter proves the importance of selecting the best model for the mid-IR data to derive meaningful quantities.} Esta an sección6.2: We suggest for future works to test as many models as possible with multi-wavelength spectroscopy to try to disentangle which models better reproduces the data
Furthermore, the $\rm{\sigma}$ parameter could be related to the opening angle of the ionization cone, which is a tracer of [OIII] emission \citep[e.g.][]{Garcia-Bernete19}. \citet{Schmitt03} presented the observation in the [OIII] filter from \emph{Hubble Space Telescope} of IC\,5063. They found that this emission is extended and aligned with the radio emission and the host galaxy major axis. These results are similar to those found by \citet{Morganti98}. According to \citet{Schmitt03} the [OIII] emission can be represented by a bicone centered at the nucleus, with an opening angle of $\rm{\alpha(cone) = 60^{\circ}}$, extending for $\rm{\sim}$2,6\,kpc along $\rm{P.A.=-65^{\circ}}$ and $\sim$ 660\,pc along the perpendicular direction. Using this measurement of the ionization opening angle, we obtained a free of cone half opening angle of $\alpha(cone-free) \sim 60^{\circ}$ \footnote{$\rm{\alpha(cone-free) = \frac{180^{\circ} - \alpha(cone)}{2}}$}. This suggests the torus occupies a free of cone area although it does not fill it up completely.

\begin{figure*}
    \centering
    \includegraphics[width=1.\columnwidth]{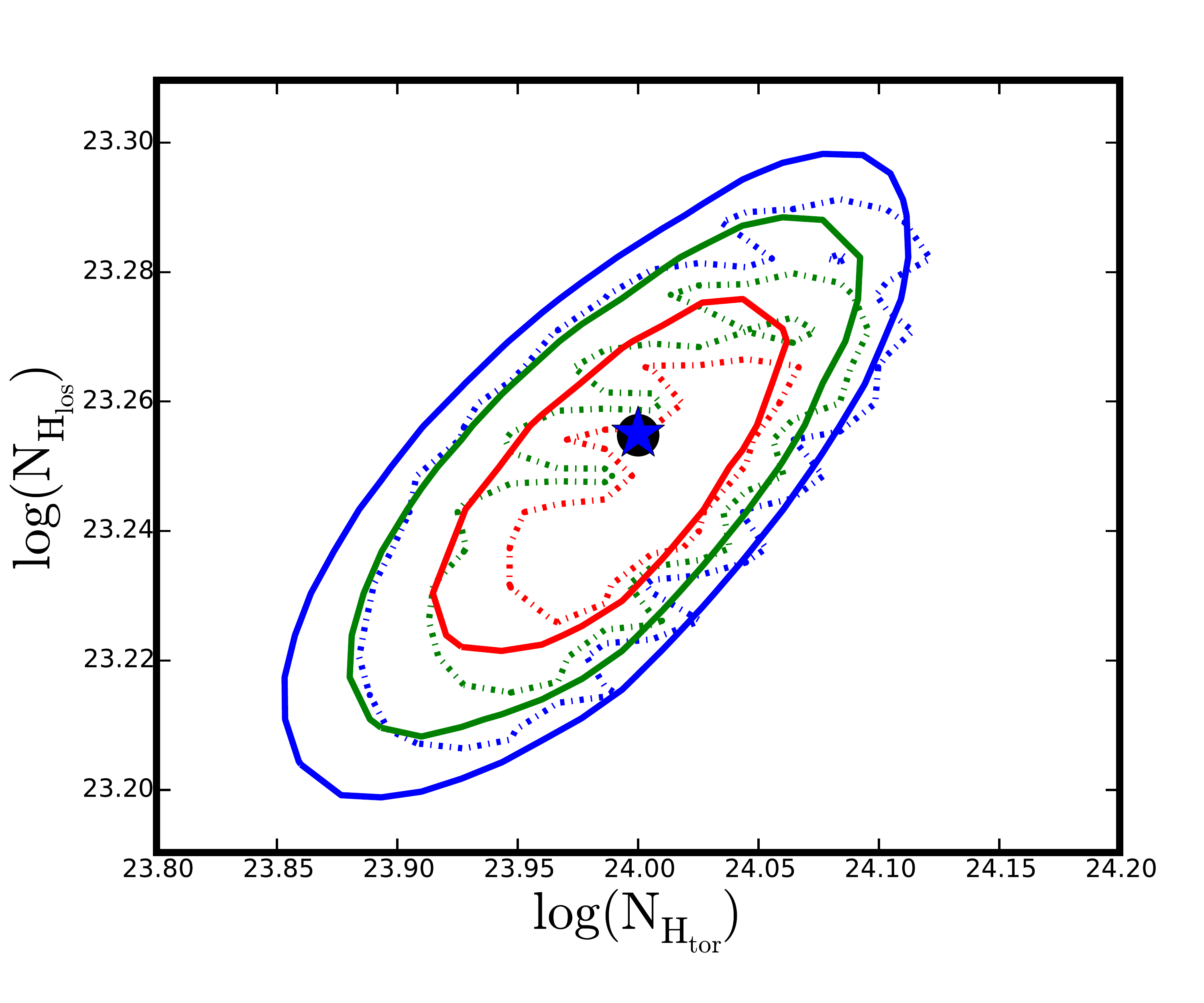}
    \includegraphics[width=1.\columnwidth]{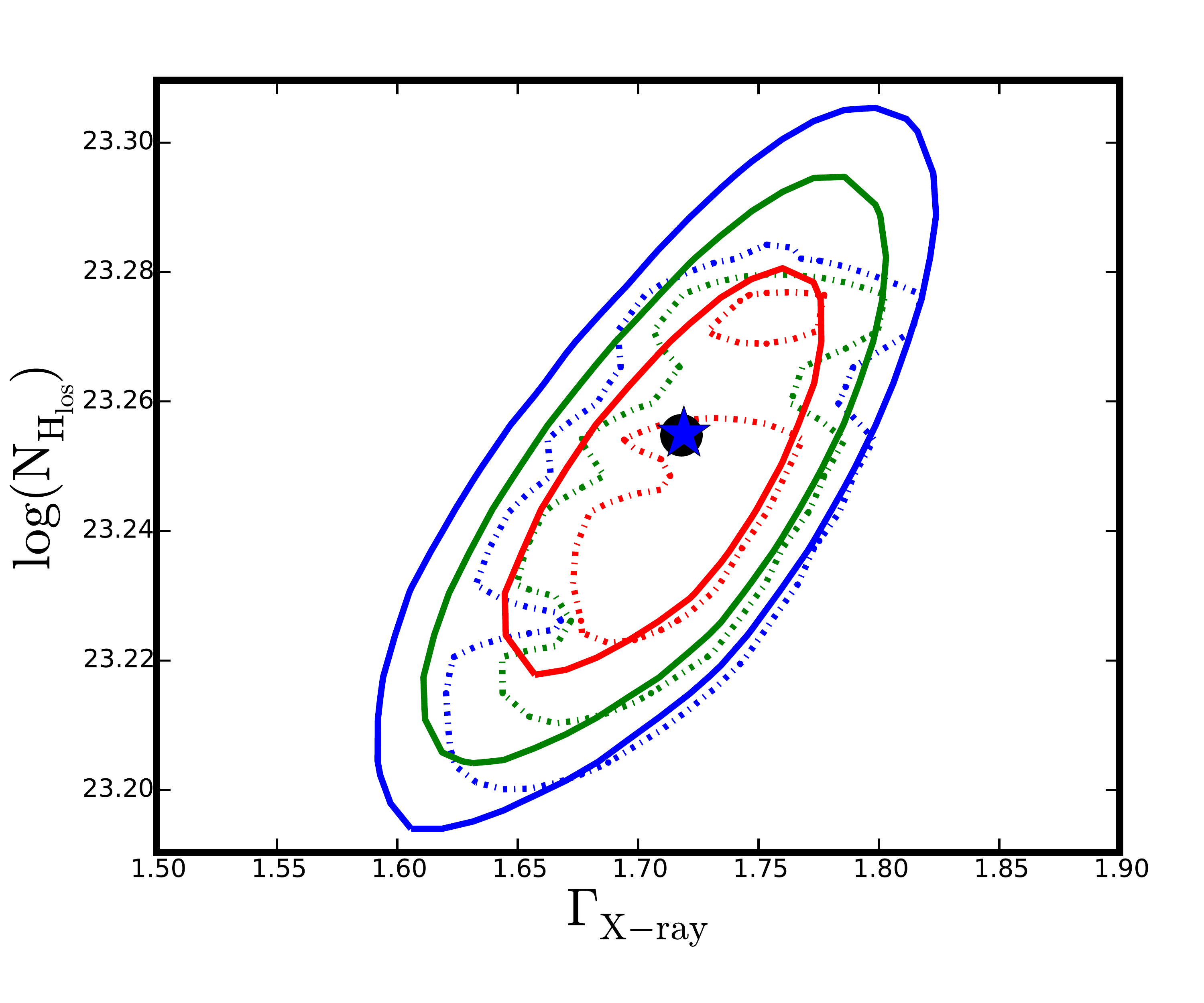}
\caption{Two dimensional $\rm{\Delta_{\chi^2}}$ contours for the LOS, density profile, and the torus $\rm{N_{H_{tor}}}$ for bS (dotted lines) and bS2 (solid lines) baseline models. The red, green, and blue lines are 1, 2, and $3\,\rm{\sigma}$ contours. The values reported in table \ref{Table:Individual_fit} are shows with a black circle and blue star for bS and bS2 baseline models, respectively.}
    \label{fig:NhtorvsNhlos}
\end{figure*}
%%at X-rays is capable of separating the 
%obscuring the line of sight
The borus02 model is capable of separating the density of the reprocessed material and that of the LOS. This option allows us to explore if the material that produces the reflection component is different from that producing the obscuration along the LOS \citep[see also][]{Balokovic18}. We tested the scenario in which both $\rm{N_H}$ are linked. We found that $\rm{\Delta \chi^2}$ increases ($\rm{\chi^2 / d.o.f = 706.15/649}$) if we fit the IC\,5063 spectra using the bS2 baseline model and we assume that $\rm{N_{H_{los}} = N_{H_{tor}}}$ (best fit result in $\rm{log(N_H)=23.25 \pm 0.02}$). We compared this $\rm{\chi^2/d.o.f.}$ with bS2 statistic through f-test and we obtained a low probability value of $\rm{1.91 \times 10^{-6}}$ for the null hypothesis. Therefore, a scenario where these two values are different is statistically preferred. In Figure \ref{fig:NhtorvsNhlos} (left), we show the two dimensional $\chi^2$ distributions for $\rm{N_{H_{los}}}$ versus $\rm{N_{H_{tor}}}$ parameters when we used the bS2 baseline model. We found that these parameters are not the same at $3\sigma$ level. Therefore, $\rm{N_{H_{tor}}}$ is larger than the $\rm{N_{H_{los}}}$ beyond parameter degeneracy (see also Table \ref{Table:Individual_fit}). Note that in a scenario where both parameters belong to the same structure it is not feasible that $\rm{N_{H_{tor}}}$ is larger than $\rm{N_{H_{los}}}$, as it is our case (see Table \ref{Table:Individual_fit}), because the former is the average density and the latter decreases with the azimuthal angle. Therefore, $\rm{N_{H_{los}}}$ and $\rm{N_{H_{tor}}}$ parameters measure the $\rm{N_H}$ of different absorbing materials. This result is also found by \citet{Balokovic18}. As a final caveat on the subject, we found that $\rm{N_{H_{los}}}$, $\rm{N_{H_{tor}}}$, and $\rm{\Gamma}$ parameter are partially degenerated. We think that this coupling between parameters is due to the natural degeneracy between obscuration and power-law steepness, where high $\rm{N_{H_{los}}}$ and low $\Gamma$ could mimic, at a certain level, to low $\rm{N_{H_{los}}}$ and high $\rm{\Gamma}$ (see Figures \ref{fig:NhtorvsNhlos} and \ref{fig:ContoursFritzfree}).

Another way to explore the properties of the torus taking into account the information at both wavelength ranges is through the column density and the optical depth from X-ray and mid-IR models, respectively. The two parameters are associated with the density of the obscuring material. The link between these parameters is not straightforward because the $\rm{N_{H_{tor}}}$ is an average measurement of the column density at the inner parts of the torus (where the reflection is produced), while the $\rm{\tau_{9.7}}$ is a measurement of the equatorial optical depth. We considered the relationship between extinction and column density, assuming a constant dust-to-gas ratio and the relation between the optical depth at 9.7\,$\rm{\mu m}$ and that in V band \citep{NenkovaB08,Feltre12} \footnote{$\rm{\tau_{9.7} = 0.042*\tau_{\nu}}$}. Following these considerations and using the values reported in Table \ref{Table:Individual_fit} for the bS2 baseline model, we obtained a column density of $\rm{log(N_{H_{\tau_{9.7}}}) > 23.65 \, cm^{-2}}$ in the equatorial plane. This value is already consistent with $\rm{N_{H_{tor}}}$. However, strictly speaking, to compare it with $\rm{N_{H_{tor}}}$, we must calculate the average column density using the dependence of the density distribution with the azimuthal angle. Nevertheless, the resulting upper limit will always be less restrictive that the $\rm{N_{H_{\tau_{9.7}}}}$ reported above.
On the other hand, we also compare the expected LOS column density from the dust distribution with the actual calculated the LOS column density $\rm{N_{H_{los}}}$ considering the inclination angle and the dust density distribution values from the bS2 baseline model\footnote{We used the following equation to calculate the $\rm{N_H}$ in the sight of line derived from the mid-IR optical depth:
    $N_{H_{\tau_{9.7}}}^{inc}(cm^{-2}) = \frac{1.086}{0.042}*\tau_{9.7}*1.9 \times 10^{21}*e^{-\gamma cos (\theta_{inc})}$. This equation consider the density function \citep[see equation 3 in][]{Fritz06}, which depend on $\theta_{inc}$ and $\gamma$.}.
The resulting value is $\rm{log(N_{H_{\tau_{9.7}}}^{los}) > 23.03}$, which is consistent with the $\rm{N_{H_{los}}}$ value. %Therefore, we produce a consistent picture explaining at the same time mid-IR and X-ray spectra, where the same viewing and half-opening angles reproduce both component. Moreover, the (average and LOS) column densities at X-ray are also consistent with optical depth of the same structure. 

Therefore, we were able to find evidence suggesting that the structure that produces the continuum (mid-IR) and the reprocessed (X-ray) emissions is the same. This suggests that the reflection component has its origin in the AGN torus.
Indeed, the smooth and the borus02 models individually fitting the mid-IR and X-ray spectra (individual fits reported in Table \ref{tab:Onlyborus02_smoothfit}), also show consistent viewing angles and width of the torus, although the $\theta_{tor}$ and $\theta_{inc}$ parameters are better restricted when using the bs2 baseline model.
This result has been largely argued in literature although we lacked of observational evidence. Although without simultaneous fitting, \citet{Farrah16} also look for the similarities on the geometrical distribution resulting from the mid-infrared and X-ray spectroscopic analysis of the radio galaxy IRAS 09104+4109. They concluded that both obscurers are consistent with being co-aligned, although viewing angle needed to be fixed to that obtained at mid-infrared wavelengths. \citet{Bianchi07} suggested that the widespread presence of a Compton reflection component strongly favors a scenario where most of the FeK$\rm{\alpha}$ emission comes from the torus and \citet{Bianchi12} listed reasons why these two components come from a region smaller than 100\,pc, associating it to the AGN. One of the strongest arguments in favour of an origin of the reflection component on the torus comes from the FeK$\rm{\alpha}$ emission line, always attached to the Compton-hump. \citet{Iwasawa93} reported for the first time an anti-correlation between the strength of the neutral narrow core of the FeK$\rm{\alpha}$ emission line and the 2-10\,keV luminosity (the so-called ``X-ray Baldwin effect'' or ``Iwasawa-Taniguchi'' effect \footnote{The baldwin effect is an anti-correlation between the equivalent width and the luminosity found in optical/UV lines.}). \citet{Page04} have also explored this effect and suggested that a possible explanation is a decrease in the \textit{Cf} of the Compton thick torus when the luminosity increases \citep[see also][]{Boorman18}. The current work shows one of the first direct evidences of the link between the reflection component and the torus.

%Therefore, we could be demonstrating for the first time that the reflection component has its origin in the torus for a source.}
%%for the first time we demonstrate 
%{\bf OGM: Aqui lo que falta es buscar paper de rayos X donde digan que esta emision debe provenir del toroide (explicando por que, principalmente por la linea de hierro que se ve estrecha) y decir que nosotros hemos sido los primeros en demostrarlo para un objeto. }

%The arguments given up to this point, about how to link parameters of mid-IR and X-ray models and their implications, take us to the conclusion that the structure that produce the continuum (mid-IR) and the reprocessed (X-ray) emissions is the same.
\begin{table}[]
    \centering
    \begin{tabular}{c|c}
        \hline
        Parameter & Value  \\
        \hline \hline
        $\Gamma$ & 1.73 $\pm_{0.07}^{0.01}$ \\
        $log(N_{H_{tor}})$ & 24.00 $\pm_{0.04}^{0.04}$ \\
        $log(N_{H_{los}})$ & 23.25 $\pm_{0.01}^{0.03}$ \\
        $\theta_{tor}$ & 59.9 $\pm_{15.4}^{2.5}$ \\
        $\theta_{inc}$ & 75.5 $\pm_{2.1}^{2.6}$ \\
        \hline
        $\chi^2 /$d.o.f. & $530/464$ \\
        \hline
        $i_{F06}\, (90. - \theta_{inc})$ & 14.9 $\pm_{0.4}^{0.4}$ \\
        $\sigma$ & 34.2 $\pm_{2.2}^{2.5}$ \\
        $\beta$ & 0.0 \\
        $Y$ & 14.1 $\pm_{0.3}^{0.3}$  \\
        $\tau_{9.7}$ & $>$ 8.9\\
        $\gamma$ & $>$ 5.6 \\
        \hline
        $\chi^2 /$d.o.f. & $148/181$ \\
        \hline \hline
    \end{tabular}
    \caption{The best-fit physical parameters using the borus02 (top) and the smooth (bottom) models for IC\,5063.}
    \label{tab:Onlyborus02_smoothfit}
\end{table}

\subsection{Parameters and derived quantities of the dusty structure}\label{sec:discus2}

%\textbf{The $Y$ and $\theta_{inc}$ are similar and $\sigma$ value is twice   to those obtained for the bS2 baseline model, while the $\sigma$ value is twice our value. Note that the $\sigma$ value is twice our value for this parameter, while the other two parameters are similar.}
%The $Y$ and $\theta_{inc}$ are similar to those obtained for the bS2 baseline model. Note that the $\sigma$ value is twice our value for this parameter, while the other two parameters are similar. {\bf OGM: ests ultima frase no la entiendo}.

The dust torus parameters of IC\,5063 have been explored using the Clumpy model and a Bayesian approach on high-resolution spectra and/or photometry by \citet{Ramos-Almeida11, Alonso-Herrero11}. The values obtained using our bC1 baseline model, except for $\theta_{inc}$ and $q$ do not agree with theirs. However, it should also be noted that the statistic obtained for this baseline model is not the best for this source. If we compare their values with the ones from our best baseline model (i.e. bS2) we find the $Y$ and $\theta_{inc}$ are in well agreement with their error ranges. Meanwhile, our $\sigma$ value is half of their reported value.
%\citet{Ramos-Almeida11} used the Clumpy model and a Bayesian approach to fit the near- and mid-IR high-resolution photometry from T-ReCS/Gemini. They obtained a $\sigma \, \sim 51$, $Y \sim 12$, $N0 \, \sim 13$, $q \sim 0.8$, $\tau_{\nu} \sim 90$, and $\theta_{inc} \, \sim 81$ for IC\,5063. The values obtained using our bC baseline model, except for $\theta_{inc}$ do not agree with these values. \citet{Alonso-Herrero11} used the same model to explore the dust torus, but they take into account both photometry and spectroscopic data from near to mid-IR wavelengths including also a foreground dust screen. They did not restrict any of the Clumpy model parameters for this object. However, they reported the median values of parameters obtained: $\sigma \, \sim 60$, $Y \sim 13$, $N0 \, \sim 14$, $q \sim \, 2.6$, $\tau_{\nu} \sim 130$, and $\theta_{inc} \, \sim 82$ for this source. Only the $\theta_{inc}$ and $q$ values are similar to those obtained in our analysis using the bC baseline model. Note also that statistic obtained using bC baseline model is not the best for this source. However, if we compared these values with those obtained for the best baseline model (i.e. bS2), the $Y$ and $\theta_{inc}$ are similar while the $\sigma$ value is half of the value reported in previous works.
Indeed, the smooth model accounts for the same dust in a smaller volume compared to the clumpy models \citep[see also][]{Gonzalez-Martin19a,Gonzalez-Martin19b}. As suggested in the literature, the resulting model parameters and derived quantities seem to strongly depend on the model used, wavelength and/or kind of data (i.e. spectroscopy or photometry). While they mostly rely on near- and mid-IR photometry (with ground-based Q-band spectroscopy), we use spectroscopic data covering mid-IR and X-ray.

%%%% Remove by dca Sept 18: On the other hand, at X-ray \citet{Tazaki11} fitted the \emph{Suzaku} and \emph{Swift}/BAT spectra of IC\,5063 using {\sc pexrav} as the reflection model. We found that our values from bS2 baseline model are the same within the error ranges reported by them. In fact, the mid-IR viewing angles found in previous works are in good agreement. However, note that the error ranges for the parameters reported in those works are not completely constrained.
%%% Escrito por OGM y DCA: \citet{Balokovic18} also reported the $\Gamma = 1.75$, $log(N_{H_{tor}}) = 23.3 $, $log(N_{H_{los}}) = 23.9$, and $\theta_{inc} > 52^{\circ}$, which are consistent with our values within 1$\sigma$ for the individual fits (see Table \ref{tab:Onlyborus02_smoothfit}) and those obtained for the bS2 baseline models (see Table \ref{Table:Individual_fit}).
On the other hand, \citet{Balokovic18} fitted the \emph{NuSTAR} spectra of IC\,5063 using borus02 model. Their values of $\Gamma = 1.75$, $log(N_{H_{tor}}) = 23.3 $, $log(N_{H_{los}}) = 23.9$, and $\theta_{inc} > 52^{\circ}$ are consistent with our values within 1$\sigma$ for the individual fits (see Table \ref{tab:Onlyborus02_smoothfit}) and those obtained for the bS2 baseline models (see Table \ref{Table:Individual_fit}). The largest discrepancy is found for the $\theta_{tor}$ parameter. \citet{Balokovic18} found $\theta_{tor} < 40^{\circ}$ which is consistent with the reported value using the individual fit and the bS1 baseline model (see Tables \ref{tab:Onlyborus02_smoothfit} and \ref{Table:Individual_fit}) at the 3$\sigma$ (see Figure \ref{fig:ContoursFritzfree}). However, this value is not consistent with that obtained by the bS2 baseline model. Note that these values are obtained assuming fix $N_{H_{tor}}$. This might explain the discrepancy found and shows the difficulties to restrict the $\theta_{tor}$ using X-ray data alone, reinforcing the need to produce a consistent picture using multiwavelength information.

This issue is also visible in the derived quantities. In the last row of Table \ref{Table:Individual_fit} we report the dust masses obtained from the mid-IR parameters, which cover a range $\rm{0.06-30.3 \, \times \, 10^5 \, M_{\odot}}$. Therefore, the dust mass depends on the chosen baseline model to fit the spectra. Despite this, the values are consistent with the ranges reported in other works \citep{Mor09, Fritz06}. Furthermore, \textit{Cf} is strongly dependent on the baseline model used. \citet{Ramos-Almeida11} compared the properties of a large sample of Sy1 and Sy2 Seyfert tori using the clumpy torus models. They found that the dusty torus in Sy2 is wider than in Sy1, and is composed of a larger number of clouds with lower optical depth. Our mid-IR covering factors ($Cf_{midIR}$ in Table \ref{Table:Individual_fit}) are consistent with their results \citep[see also][]{Brightman15}.

We also calculated the $\rm{R_{out}}$ for each of our baseline models. We find $R_{out} > 23.9$\,pc (diameter $\rm{\sim}$ 0.2") and $\rm{R_{out} = 3.4}$\,pc (diameter $\rm{\sim}$ 0.03") when using the bC1 and bS2 baseline models to fit the spectra, respectively. This last $R_{out}$ value is consistent with the reported by \citet{Ichikawa17} for IC\,5063. Therefore, the selection of the baseline model is crucial to obtain meaningful results for both structural parameters and derived quantities. Additionally, note that only the $R_{out}$ obtained using the bC1 baseline model could be detectable at the best spatial resolution provided by ALMA\footnote{The highest spatial resolution obtained with ALMA uses a configuration C43-10 in band 7}. However, ALMA data is also sensitive to the radio jet emission, so a proper study of the SED is required to use ALMA data to study the AGN dust \citep{Pasetto19}. Finally, we suggest for future works to test as many models as possible with multi-wavelength spectroscopy to try to disentangle which models better reproduces the data before drawing any conclusion on the parameters.    

Overall, according to the values of parameters results using the bS2 baseline model, the IC\,5063 torus is compact ($R_{out} \sim 3.4$\,pc) and relatively thin ($\sigma \sim 34$) structure. Our bS2 baseline model solution also favors a dust torus in which the density profile only has an azimuthal dependence ($\gamma > 5.6$), i.e. a strong decrease on the dust/gas density when the half-opening angle increases. %\textbf{We found that the density of the torus calculated from the $\tau_9.7$ parameter is $\rm{>23.65\,cm^{-2}}$, which it is consistent with the $NH_{tor}$ value.}

\subsection{About our simultaneous fitting technique}\label{sec:discus3}
%Despite producing an overall explanation for both 
Apart from simultaneously explaining both mid-IR and X-ray continuum emission, the main advantage of being able to link some parameters from mid-IR and X-ray models is that we can find all of them from the final fit (see Fig. \ref{fig:ContoursFritzfree}). Therefore, we can obtain more information and explore the source of obscuration at both wavelengths.

In the case of IC\,5063, we found that the best option to fit its spectra is using a combination of the Smooth and borus02 models (bS2). A caveat on this result is that these two models may have been our best choice to fit the data due to their geometric similarities. Indeed, both models assume a smooth distribution arranged in a torus-like structure. Nevertheless, these models assume a different density distribution; the borus02 model assumes a uniform density profile for the gas distribution and the smooth model considers that it decreases towards large azimuths and radii. Additionally, the smooth model assumes that a dusty structure is located between an inner and outer radius while gas can reach the accretion disc.
%\textbf{Finally, note that the mid-IR model most similar to borus02 is the smooth model. There are, however, several differences. Firstly, the smooth model assumes that a dusty structure is located between an inner and outer radius. Additionally, it considers the distribution of material decreases towards the external parts (i.e., change with the increase of radius and half opening angle). The borus02 model is simpler because it assumes a uniform density -i.e., it does not consider the changes in the distribution of material toward inner and external parts of the torus.}
Recently, \citet{Tanimoto19} constructed the XCLUMPY model that is the radiative transfer of neutral gas at X-rays using the same distribution as the clumpy torus at mid-IR. A combination of XCLUMPY and Clumpy models might also get good results. We can discard this scenario for IC\,5063 since the residuals seen at mid-IR for the Clumpy model are significantly larger than those reported for the smooth mode. 
%Unfortunately, this model is not publicly available by the time of this publication.
However, as we expect different AGN to be better reproduced with different models, we will explore this possibility using a AGN sample in a forthcoming paper (Esparza-Arredondo in preparation).

Finally, a few words on the applicability of this technique to AGN samples. Our technique of simultaneous fitting can be applied to any type of AGN that is not dominated by the host galaxy. The best results could be found when using the high-spatial resolution mid-IR spectra and hard ($\rm{>10 keV}$) X-ray spectra to ensure a proper decontamination of the host galaxy and a characterization of the reflection component, respectively. The reflection dominated spectra at X-ray (i.e. with high obscuration towards the LOS) are also better targets. \emph{Spitzer}/IRS spectra can be used as long as the AGN dominates the emission, future \emph{JWST} observations would be needed otherwise.

% OGM: se puede aplicar esta tecnica a cualquier agn? que tipo de objectos necesitas wue sean? que dice balokovic de como deben ser los esperros para derivar los psrametros mehor? necesitas alta resolucion espacial en el mir? }

\section{Conclusions}\label{sec:Conclusion}
In this work, we explored if the X-ray reflection component and the mid-IR continuum of AGN are linked to the same structure, i.e. the so-called AGN torus. Showing that is the case, we also investigate if the combination of X-ray and mid-IR spectra and different torus models could help us to restrict the torus physical parameters of the nearby Seyfert IC\,5063 galaxy. We considered \emph{Spitzer}/IRS and \emph{NuSTAR} spectra for this analysis. We combined the radiative transfer code borus02 at X-ray \citep{Balokovic18} to describe the X-ray reflection and Smooth \citep{Fritz06}, Clumpy \citep{NenkovaB08}, or CAT3D-wind \citep{Hoenig17} models to describe mid-IR AGN dust to create a set of baseline models.
We found that the combination of the borus02 and Smooth models is the best choice to fit the spectra from both wavelengths of IC\,5063. Moreover, all the parameters of the dusty torus can be constrained if the X-ray and mid-IR inclination and half-opening angles are linked to the same value (bS2 baseline model). This link between parameters suggests that the same structure producing the reflection component is emitting through dust heating at mid-IR. This could be the first time such behavior is confirmed by comparing the expected morphology at and obscuring material distributions at both wavelengths. This technique can be used to infer the physical properties of the torus of any AGN that is not dominated by the host galaxy at mid-IR and shows a significant fraction of the reflection component at X-ray. 

%\textbf{We found that the combination of the borus02 and Smooth models is the best choice to fit the mid-IR and X-ray spectra of IC\,5063. Indeed both the inclination angle and the angular width of the torus can be linked to the same value indicating that the same structure that produces the reflection component is emitting through dust heating at mid-IR.} This is the first time such behavior is confirmed. Moreover, we found that all the parameters of the dusty torus are found when the inclination and half-opening angles are linked to vary together among both baseline models. Our main conclusion is that this technique can be used to infer the physical properties of the torus of any AGN that is not dominated by the host galaxy at mid-IR and shows a significant fraction of the reflection component at X-rays. We will explore the properties of the torus using a sample of AGN in a \textbf{forthcoming} analysis. 
%We found that the combination of the borus02 and Smooth models is the best choice to fit the mid-IR and X-ray spectra of IC\,5063. Indeed both the inclination angle and the angular width of the torus can be linked to the same value indicating that the same structure that produces the reflection component is emitting through dust heating at mid-IR.
%We combined the radiative transfer codes from both wavelengths to create a baseline models.
%We used the high-resolution IRS/\emph{Spitzer} and \emph{NuSTAR} spectra of the nearby type-2 IC\,5063 Seyfert galaxy.

\acknowledgments
%The authors thank the anonymous referee for careful reading and constructive suggestion that improved the paper.
The authors thank the anonymous referee for careful reading and constructive suggestion that improved the paper.
This work made use of data from the NuSTAR mission, a project led by CalTech, managed by JPL, and funded by NASA. We thank the NuSTAR Operations, Software and Calibration teams for support with the execution and analysis of these observations. This research has made use of the NuSTAR Data Analysis Software (NuSTARDAS) jointly developed by the ASI Science Data Center (ASDC, Italy) and CalTech. This work is based in part on observations made with the \emph{Spitzer} Space Telescope, which is operated by the Jet Propulsion Laboratory, California Institute of Technology under a contract with NASA.
D. E.-A. acknowledges support from a CONACYT scholarship.
This research is mainly funded by   the UNAM PAPIIT project IA103118 (PI OG-M). MM-P acknowledges support by KASI postdoctoral fellowships. 
JM acknowledges financial support from the research project AYA2016-76682-C3-1-P (AEI/FEDER, UE) and the State Agency for Research of the Spanish MCIU through the Center of Excellence Severo Ochoa award for the Instituto de Astrof\'isica de Andaluc\'ia (SEV-2017-0709).
C.R.-A. acknowledges the Ram\'on y Cajal Program of the Spanish Ministry of Economy and Competitiveness through project RYC-2014-15779 and the Spanish Plan Nacional de Astronom\'ia y Astrofis\'ica under grant AYA2016-76682-C3-2-P.

\end{document}

%%%%%%%%%%%%%%%%%%%%%%%%%%%%%%%%%%%%%%%%%%%%